\documentclass[11pt, oneside]{article}   	% use "amsart" instead of "article" for AMSLaTeX format
\usepackage[T1]{fontenc}
\usepackage{lmodern}
\usepackage{geometry}                		% See geometry.pdf to learn the layout options. There are lots.
\usepackage{graphicx}				% Use pdf, png, jpg, or eps§ with pdflatex; use eps in DVI mode
\usepackage{amsmath}		
\usepackage{amssymb,amsfonts,amsthm}
\usepackage{mathtools,cool}
\usepackage{braket}

\usepackage[cal=boondoxo]{mathalfa}

\title{\textbf{\Large Spectrum, Scattering Amplitudes, and String Field Theory of 4 Dimensional Twistor String}}
\author{Christian Kunz \\ \small{\textit{E-mail:} \href{mailto:kunz.christian.321@gmail.com}{kunz.christian.321@gmail.com}}}
\usepackage{plain}
\usepackage{hyperref}

\newcommand{\ud}{\mathrm{d}}
\numberwithin{equation}{section}
\allowdisplaybreaks

\begin{document}
  \maketitle
  \tableofcontents
  
\begin{abstract}
  The recently introduced anomaly-free twistor string in 4 dimensions is further explored. The spectrum based on the physical states and its Minkowski interpretation are examined. Scattering amplitudes with vertex operators involving gravitons and fermions are computed and are compared with Einstein-Yang-Mills amplitudes. Tree and one-loop scattering amplitudes are shown to have proper unitary factorization properties. Finally a string field theory is proposed.
\end{abstract}

\section{Introduction}
     Recently a 4 dimensional anomaly-free twistor string has been presented \cite{Kunz:2020}. The target space is twistor space with its dual, the string contains a target bi-twistor and two target fermionic bi-spinors, all of them being worldsheet spinors of conformal weight $\frac{1}{2}$, and besides being coupled to two-dimensional worldsheet gravity, the model is gauged with regards to a rotational SU(2) symmetry between the two twistors. Additionally the conformal translations between the twistors and between twistors and fermions are gauged as well. It was shown that this string is anomaly-free with nilpotent BRST charge and reproduces the familiar N$^k$MHV amplitudes of N=8 supergravity \cite{Skinner:2013,Geyer:2014}.
          
     In this follow-up paper the properties of the string are explored in more detail. We limit ourselves entirely to the closed left-moving string because it already provides all the interesting features. The spectrum of the string is analyzed with help of vertex operators for all known physical states and is shown to have rich variety. There are four gravitons. One of them leads to expected gravitational amplitudes, the others do not. The remainder of the spectrum exhibits an SU(4) target R symmetry from the fermionic spinors and an SU(2) symmetry from the bi-twistors. What is particularly interesting is that with a judicious choice of an SU(2) representation, it includes three generations of fermionic content of the Pati-Salam model \cite{Pati:1974}. Scattering amplitudes involving vertex operators with gravitons and fermions look like the Einstein-Yang-Mills (EYM) amplitudes of \cite{Adamo:2015}. These new amplitudes and also the tree and one-loop scattering amplitudes computed in \cite{Kunz:2020} exhibit expected unitary factorization. Finally, a string field theory along the lines of \cite {Lacroix:2017} and \cite{Reid:2017} is proposed.
          
     In appendix \ref{StringProperties} we list some basic properties of the string, including vacuum normalization, picture changing operators (PCO), and a list of physical states.

     In section 2 we look at fixed vertex operators which produce the physical states identified in appendix \ref{StringProperties} and are directly related to Penrose transforms. The Neveu-Schwarz (NS) sector has two gravitons, chosen to be in an SU(2) singlet representation. One of them reproduces the proper gravitational amplitudes, but the other one does not. For the latter a question is raised whether it actually represents a massless particle. The NS sector also contains 16 spin $\frac{3}{2}$ fermions and 22 vector bosons in SU(4) representations of the R symmetry. There are neither spin $\frac{1}{2}$ fermions nor scalars in the NS sector. The spectrum in the Ramond (R) sector is much richer and contains two gravitons and particles of spin $\frac{3}{2}, 1, \frac{1}{2},$ and 0 in SU(4) representations. The gravitons are again taken to be in an SU(2) singlet representation, but for the other particles the SU(2) representation is left open. One interesting observation is that there is a 24-dimensional SU(4) representation for spin $\frac{1}{2}$ fermions that, when chosen to be in a fundamental SU(2) representation, makes up an SU(2) doublet of six fundamental SU(4) representations, reminiscent of the Pati-Salam model \cite{Pati:1974}. The gravitons in the R sector lead to improper gravitational amplitudes.
         
     In section 3 we find a few integrated vertex operators for external gravitons and fermions, and compute tree scattering amplitudes in the NS sector that are shown to match the single-trace EYM amplitudes of \cite{Adamo:2015} for certain collections of fermions.
     
     In section 4 we investigate the factorization of the scattering amplitudes of section 3 and the ones computed in \cite{Kunz:2020}. This is done by modeling separating and non-separating degenerations of the worldsheet. The resulting amplitude can then be interpreted as containing one or two additional particles propagating on-shell with proper factorization. In particular, it is shown explicitly that the non-separating factorization of the four-graviton one-loop amplitude leads to the expected result.
     
     In section 5 a string field theory is sketched out taking clues from \cite{Reid:2017} and \cite{Lacroix:2017}. The classical limit, if it exists, shows resemblance to twistor actions proposed in the literature \cite{Adamo_2:2013}.

     The last section contains summary and outlook.
     
     In the following sections the notation of appendix \ref{StringProperties} for twistor, fermionic, ghost, and anti-ghost fields and for the BRST charge and current components will be adopted everywhere.\\

\section{Vertex Operators and Spectrum}
  \label{Spectrum}
    
  We define fixed vertex operators for every physical state listed at the end of appendix \ref{StringProperties}. They are in ghost number 1 of every fermionic ghost field and picture number $-1$ for every bosonic ghost, and, therefore, are all multiplied with the ghost factor
  \begin{equation}
  \mathcal{G\!F}  =  c \!\prod_{r=1}^3 g^r \!\!\! \prod_{\alpha, \dot{\alpha}=1}^2 \!\!\! \left(\!\prod_{i,j=1}^2 \!\! e_{i j}^{\dot{\alpha} \alpha} \! \prod_{i,j,l=1}^2 \!\! \delta(\gamma_{i j l}^{\alpha \dot{\alpha}})\!\!\right). 
  \label{GhostFactor}
  \end{equation}
  First we extend the vertex operators of \cite{Kunz:2020} to include fermions. A typical fixed vertex operator for the NS sector is
  \begin{equation}
  \begin{split}
  \mathcal{V}_{i} \!& = \mathcal{G\!F}
   \int \! \frac{\ud t} {t^3}\! : \!\prod_{s=1}^2 \!\left(\! \delta^2\!(\rho_i \!-\! t \lambda_s(z))\,  \delta^2\!(\eta_{is} \!-\! t \phi_s(z)) \!\right)\! \mathrm{e}^{it  \sum_{s=1}^2( [\tilde{\rho}_i \mu_s(z)] + [\tilde{\eta}_{is} \psi_s(z)]) }\! \!: \,, \\
  \tilde{\mathcal{V}}_{i} \!& = \mathcal{G\!F}
  \int \! \frac{\ud t} {t^3} \! : \! \prod_{s=1}^2 \!\left(\! \delta^2\!(\tilde{\rho}_i \!-\! t \tilde{\lambda}_s(z)) \,\delta^2\!(\tilde{\eta}_{is} \!-\! t \tilde{\phi}_s(z)) \!\right)\!  \mathrm{e}^{it \sum_{s=1}^2(\braket{\tilde{\mu}_s(z) \rho_i} + \braket{\tilde{\psi}_s(z) \eta_{is}}) }\!\! : \,,
  \end{split}
  \label{FixedVertex}
  \end{equation}
  where we used the notation $\!\braket{\rho \tilde{\mu}}\! = \rho_{\alpha} \tilde{\mu}^{\alpha} = \epsilon_{\beta \alpha} \rho^{\beta} \tilde{\mu}^{\alpha} , [\tilde{\rho} \mu] = \rho_{\dot{\alpha}} \mu^{\dot{\alpha}} = \epsilon_{\dot{\beta} \dot{\alpha}} \rho^{\dot{\beta}} \mu^{\dot{\alpha}}$.
  
  \eqref{FixedVertex} describes a plane wave for a massless particle with momentum $p_i^{\alpha \dot{\alpha}} \sim \rho_i^{\alpha} \tilde{\rho}_i^{\dot{\alpha}}$. The $\eta_{is}$ and $\tilde{\eta}_{is}$ are Grassmann odd variables similar to the ones used for supermomenta in supersymmetric models and are used here for convenience. These vertex operators are BRST-closed. Actually, using the operator product expansion one ends up with terms proportional to  
  \begin{equation}
  \begin{aligned}
  &(e_{i j}^{\dot{\alpha} \alpha} \partial e_{i j}^{\dot{\alpha} \alpha})(z)\,,\,\,\,&
  &(\partial \gamma_{1 i s}^{\alpha \dot{\alpha}} \delta ( \gamma_{1 i s}^{\alpha \dot{\alpha}}))(z)\,,&
  &(\partial \gamma_{2 i s}^{\dot{\alpha} \alpha} \delta( \gamma_{2 i s}^{\dot{\alpha} \alpha}))(z)\,,\,\,\,&
  &(g^r \partial g^r)(z)\,.\,\,\,&
  \end{aligned}
  \label{anomalies}
  \end{equation}
  But these operators do not contribute to the cohomology of Q, therefore,  give zero contributions on-shell, and will be discarded, here and for all vertex operators listed from now on  \footnote{In \cite{Kunz:2020} this was done implicitly without special mentioning.}. \\
 
  Additional BRST-closed fixed vertex operators that include dual matter fields are in both the NS sector and the R sector\footnote{As the vertex operators refer to plane waves, they are in the R sector necessarily given in coordinates on the cylinder.}:
  \begin{align}
  \mathcal{V}_{ir}^{(0)} \!&=\! \mathcal{G\!F} \!\!\!
   \int \! \frac{\ud t} {t^2}\! : \! [\tilde{\rho}_i \tilde{\lambda}_r(\!z\!)] \, \prod_{s=1}^2 \!\left(\! \delta^2\!(\rho_i \!-\! t \lambda_s(z))\,  \delta^2\!(\eta_{is} \!-\! t \phi_s(z)) \!\right)\! \mathrm{e}^{it  \sum_{s=1}^2( [\tilde{\rho}_i \mu_s(z)] + [\tilde{\eta}_{is} \psi_s(z)]) }\! \!: \,,\nonumber\\
  \tilde{\mathcal{V}}_{ir}^{(0)} \!&= \!\mathcal{G\!F} \!\!
  \int \! \frac{\ud t} {t^2} \! : \! \!\braket{\rho_i \lambda_r(\!z\!)} \, \prod_{s=1}^2 \!\left(\! \delta^2\!(\tilde{\rho}_i \!-\! t \tilde{\lambda}_s(z)) \,\delta^2\!(\tilde{\eta}_{is} \!-\! t \tilde{\phi}_s(z)) \!\right)\!  \mathrm{e}^{it \sum_{s=1}^2(\braket{\tilde{\mu}_s(z) \rho_i} + \braket{\tilde{\psi}_s(z) \eta_{is}}) }\!\! : \,,
  \label{FixedVertex1}
  \end{align}
  
  \begin{align}
  \mathcal{V}_{ir}^{(1)} \!&=\! \mathcal{G\!F} \!\!\!
   \int \! \frac{\ud t} {t^2}\! : \! [\tilde{\eta}_{ir} \tilde{\phi}_r(\!z\!)] \, \prod_{s=1}^2 \!\left(\! \delta^2\!(\rho_i \!-\! t \lambda_s(z))\,  \delta^2\!(\eta_{is} \!-\! t \phi_s(z)) \!\right)\! \mathrm{e}^{it  \sum_{s=1}^2( [\tilde{\rho}_i \mu_s(z)] + [\tilde{\eta}_{is} \psi_s(z)]) }\! \!: \,,\nonumber\\
  \tilde{\mathcal{V}}_{ir}^{(1)} \!&= \!\mathcal{G\!F} \!\!\!
  \int \! \frac{\ud t} {t^2} \! : \! \!\braket{\eta_{ir} \phi_r(\!z\!)} \, \prod_{s=1}^2 \!\left(\! \delta^2\!(\tilde{\rho}_i \!-\! t \tilde{\lambda}_s(z)) \,\delta^2\!(\tilde{\eta}_{is} \!-\! t \tilde{\phi}_s(z)) \!\right)\!  \mathrm{e}^{it \sum_{s=1}^2(\braket{\tilde{\mu}_s(z) \rho_i} + \braket{\tilde{\psi}_s(z) \eta_{is}}) }\!\! : \,.
  \label{FixedVertex2}
  \end{align}
  Vertex operators exclusively in the R sector involving derivatives of matter fields are:
  \begin{equation}
  \begin{split}
  \mathcal{V}_{ir}^{(2)} \!& = \mathcal{G\!F} \!\!\!
   \int \!\! t \ud t : \! \!\braket{\lambda_r(z) \partial \lambda_r(z)} \! \prod_{s=1}^2 \!\left(\! \delta^2\!(\rho_i \!-\! t \lambda_s(z))\,  \delta^2\!(\eta_{is} \!-\! t \phi_s(z)) \!\right)\! \mathrm{e}^{it  \sum_{s=1}^2( [\tilde{\rho}_i \mu_s(z)] + [\tilde{\eta}_{is} \psi_s(z)]) }\! \!: \,, \\
  \tilde{\mathcal{V}}_{ir}^{(2)} \!& = \mathcal{G\!F} \!\!\!
  \int \!\! t \ud t : \!\! [\tilde{\lambda}_r(z) \partial \tilde{\lambda}_r(z)]  \prod_{s=1}^2 \!\left(\! \delta^2\!(\tilde{\rho}_i \!-\! t \tilde{\lambda}_s(z)) \,\delta^2\!(\tilde{\eta}_{is} \!-\! t \tilde{\phi}_s(z)) \!\right)\!  \mathrm{e}^{it \sum_{s=1}^2(\braket{\tilde{\mu}_s(z) \rho_i} + \braket{\tilde{\psi}_s(z) \eta_{is}}) }\!\! : \,,
  \end{split}
  \label{FixedVertex3}
  \end{equation}
  
  \begin{equation}
  \begin{split}
  \mathcal{V}_{ir}^{(3)} \!& = \mathcal{G\!F}\!\!\!
   \int \! \ud t : \! \!\braket{\eta_{ir} \partial \phi_r(z)} \prod_{s=1}^2 \!\left(\! \delta^2\!(\rho_i \!-\! t \lambda_s(z))\,  \delta^2\!(\eta_{is} \!-\! t \phi_s(z)) \!\right)\! \mathrm{e}^{it  \sum_{s=1}^2( [\tilde{\rho}_i \mu_s(z)] + [\tilde{\eta}_{is} \psi_s(z)]) }\! \!: \,,\\
  \tilde{\mathcal{V}}_{ir}^{(3)} \!& = \mathcal{G\!F} \!\!\!
  \int \! \ud t : \! [\tilde{\eta}_{ir} \partial \tilde{\phi}_r(z)] \, \prod_{s=1}^2 \!\left(\! \delta^2\!(\tilde{\rho}_i \!-\! t \tilde{\lambda}_s(z)) \,\delta^2\!(\tilde{\eta}_{is} \!-\! t \tilde{\phi}_s(z)) \!\right)\!  \mathrm{e}^{it \sum_{s=1}^2(\braket{\tilde{\mu}_s(z) \rho_i} + \braket{\tilde{\psi}_s(z) \eta_{is}}) }\!\! : \,.
  \end{split}
  \label{FixedVertex4}
  \end{equation}
    
  Now we are ready to examine the Minkowski interpretation of the particle content of the model, by looking at the linearized Penrose transforms of the vertex operators, with the $\mathcal{V}_{\cdots}^{(\cdot)}$s and $\tilde{\mathcal{V}}_{\cdots}^{(\cdot)}$s referring to particles with opposite helicity. As already mentioned, at the end of appendix \ref{StringProperties} physical states in the cohomology of the BRST operator have been identified. It is quite possible that they make up a complete set of physical states, although there is no proof. When listing the particle content we display representations of the R symmetry for the fermionic bi-spinors $\phi_{i \alpha_i}$ and $\tilde{\phi}_{j \dot{\alpha}_j}$ which is SU(4) for each of them. Notice that without supersymmetry there is no requirement that the states are in an adjoint representation of a super multiplet. Concerning the SU(2) symmetry between the two twistors which is also a target symmetry we assume that spin 2 particles (gravitons) are in an SU(2) singlet state. The SU(2) representation for other states are predetermined in the NS sector, but in the R sector they are left open, with one exception (see below). The contribution of the ghost factor $\mathcal{GF}$ to the physical states affecting normalization is omitted here and will be dealt with in section \ref{SFT}.\\
  
 \begin{table}[ht!]
 \begin{center}
 \begin{tabular}{||c|c|c||} 
 \hline
 Oscillators & Vertex Op & (Helicity| SU(4), SU(2)) \\ [0.5ex]
 \hline\hline
 &&\\ [-1.5ex]
 $\epsilon_1^{i j \alpha \beta} \lambda_{i \alpha \, -\!\frac{1}{2}} \lambda_{j \beta \, -\frac{1}{2}}$ & $\mathcal{V}_k$  & (2|1,1) \\ [1.5ex]
 \hline
 &&\\ [-1.0ex]
 $\epsilon_2^{i j \alpha \beta} \lambda_{i \alpha \, -\!\frac{1}{2}} \phi_{j \beta \, -\!\frac{1}{2}}$ & $\mathcal{V}_k$  & ($\frac{3}{2}|\bar{4},\bar{2}$) \\ [1.5ex]
 \hline
 &&\\ [-1.5ex]
 $\epsilon_3^{i j \alpha \dot{\beta}} \lambda_{i \alpha \, -\!\frac{1}{2}} \tilde{\phi}_{j \dot{\beta} \, -\!\frac{1}{2}}$ & $\mathcal{V}_{kj}^{(1)}$  & ($\frac{3}{2}|4,\bar{2}$) \\ [1.0ex]
 \hline
 &&\\ [-1.0ex]
 $\epsilon_4^{i j \alpha \beta} \phi_{i \alpha \, -\!\frac{1}{2}} \phi_{j\beta \, -\!\frac{1}{2}}$ & $\mathcal{V}_k$ & (1|6,1) \\ [1.5ex]
 \hline\hline
 &&\\ [-1.5ex]
 $\epsilon_5^{i j \dot{\alpha} \dot{\beta}} \tilde{\lambda}_{i \dot{\alpha} \, -\!\frac{1}{2}} \tilde{\lambda}_{j \dot{\beta} \, -\!\frac{1}{2}}$ & $\tilde{\mathcal{V}}_k$  & \!(-2|1,1)\\ [1.5ex]
 \hline
 &&\\ [-1.0ex]
 $\epsilon_6^{i j \dot{\alpha} \dot{\beta}} \tilde{\lambda}_{i \dot{\alpha} \, -\!\frac{1}{2}} \tilde{\phi}_{j \dot{\beta} \, -\!\frac{1}{2}}$ & $\tilde{\mathcal{V}}_k$ & \!(-$\frac{3}{2}$|4,2) \\ [1.5ex]
 \hline 
 &&\\ [-1.0ex]
 $\epsilon_7^{i j \dot{\alpha} \beta} \tilde{\lambda}_{i \dot{\alpha} \, -\!\frac{1}{2}} \phi_{j \beta \, -\!\frac{1}{2}}$ & $\tilde{\mathcal{V}}_{kj}^{(1)}$ & \!(-$\frac{3}{2}|\bar{4}$,2) \\ [1.5ex]
 \hline
 &&\\ [-1.5ex]
 $\epsilon_8^{i j \dot{\alpha} \dot{\beta}} \tilde{\phi}_{i \dot{\alpha} \, -\!\frac{1}{2}} \tilde{\phi}_{j \dot{\beta} \, -\!\frac{1}{2}}$ & $\tilde{\mathcal{V}}_k$ & \!(-1|6,1) \\ [1.5ex]
 \hline\hline
 &&\\ [-1.5ex]
 $\epsilon_9^{i j \alpha \dot{\beta}} \lambda_{i \alpha \, -\!\frac{1}{2}} \tilde{\lambda}_{j \dot{\beta} \, -\!\frac{1}{2}}$ & $\mathcal{V}_{kj}^{(0)}$ & (2|1,1) \\  [1.5ex]
 \hline 
 &&\\ [-1.5ex]
 $\epsilon_{10}^{i j \alpha \dot{\beta}} \phi_{i \alpha \, -\!\frac{1}{2}} \tilde{\phi}_{j  \dot{\beta} \, -\frac{1}{2}}$ & $\mathcal{V}_{kj}^{(1)}$ & (1|$15 \oplus 1$,1) \\ [1.5ex]
 \hline
 &&\\ [-1.5ex]
 $\epsilon_9^{i j \alpha \dot{\beta}} \lambda_{i \alpha \, -\!\frac{1}{2}} \tilde{\lambda}_{j \dot{\beta} \, -\!\frac{1}{2}}$ & $\tilde{\mathcal{V}}_{ki}^{(0)}$ & \!(-2|1,1) \\ [1.5ex]
 \hline 
 &&\\ [-1.5ex]
 $\epsilon_{10}^{i j \alpha \dot{\beta}} \phi_{i \alpha \, -\!\frac{1}{2}} \tilde{\phi}_{j  \dot{\beta} \, -\frac{1}{2}}$ & $\tilde{\mathcal{V}}_{ki}^{(1)}$ & \!(-1|$15 \oplus 1$,1) \\ [1.5ex]
 \hline
 \end{tabular}
 \caption{NS Sector. The last 2 rows refer to the same states as the previous 2, but for different vertex operators and helicities.}
 \label{NSSpectrumTable}
 \end{center} 
  \end{table}
 
 \begin{table}[ht!]
 \begin{center}
 \begin{tabular}{||c|c|c||} 
 \hline
 Oscillators & Vertex Op & (Helicity| SU(4), SU(2)) \\ [0.5ex] 
 \hline\hline
 &&\\
 $[\tilde{\lambda}_{j -1} f_j(\lambda_{k 0},\!\phi_{s 0})]$ & $\mathcal{V}_{ij}^{(0)}$  & $(2|1,1), \,\,\,\,\,\,\,\,\,(\frac{3}{2}|\overline{4},\cdot), \,\,\,\,\,\,\,\,\,(1|6,\cdot), \,\,\,\,\,\,\,\,\,(\frac{1}{2}|4,\cdot), \,\,\,\,\,\,\,\,\,(0|1,\cdot)$\\ 
 &&\\
 \hline
 &&\\
 $\braket{\lambda_{j -1} g_j(\lambda_{k 0},\!\phi_{s 0})}$ & $\mathcal{V}_{ij}^{(2)} $ & $(0|1,\cdot), \,\,\,\,(-\frac{1}{2}|\overline{4},\cdot), \,\,\,\,(-1|6,\cdot), \,\,\,\,(-\frac{3}{2}|4,\cdot), \,\,\,\,(-2|1,1)$ \\ 
 &&\\
 \hline
 &&\\
 $[\tilde{\phi}_{r -1} \mathcal{f}(\lambda_{k 0},\!\phi_{s 0})]$ & $\mathcal{V}_{ir}^{(1)}$ & $(\frac{3}{2}|4,\cdot), \,(1|4 \!\otimes\! \overline{4},\cdot), \,\,(\frac{1}{2}|4 \!\otimes\!  6,2), \,\,(0|4\!\otimes\!  4,\cdot), \,\,(-\frac{1}{2}, 4,\cdot)$ \\
 &&\\
 \hline
 &&\\
 $\braket{\phi_{r -1} \mathcal{g}(\lambda_{k 0},\!\phi_{s 0})}$ & $\mathcal{V}_{ir}^{(3)}$ & $(\frac{1}{2}|\overline{4},\cdot), (0|\overline{4} \!\otimes\! \overline{4},\cdot), (-\!\frac{1}{2}|\overline{4} \!\otimes\!  6,\overline{2}), (-\!1|\overline{4} \!\otimes\!  4,\cdot), (-\!\frac{3}{2}, \overline{4},\cdot)$ \\
 &&\\
 \hline\hline
 &&\\
 $\braket{\lambda_{j -1} \tilde{f}_j(\tilde{\lambda}_{k 0},\!\tilde{\phi}_{s 0})}$ & $\tilde{\mathcal{V}}_{ij}^{(0)}$ & $(-\!2|1,1), \,\,\,(-\frac{3}{2}|4,\cdot), \,\,\,\,\,(-1|6,\cdot), \,\,\,\,\,\,(-\frac{1}{2}|\overline{4},\cdot), \,\,\,\,\,\,\,(0|1,\cdot)$ \\
 &&\\
 \hline
 &&\\
 $[\tilde{\lambda}_{j -1} \tilde{g}_j(\tilde{\lambda}_{k 0},\!\tilde{\phi}_{s 0})]$ & $\tilde{\mathcal{V}}_{ij}^{(2)}$ & $(0|1,\cdot), \,\,\,\,\,\,\,\,\,(\frac{1}{2}|4,\cdot), \,\,\,\,\,\,\,\,\,(1|6,\cdot), \,\,\,\,\,\,\,\,\,(\frac{3}{2}|\overline{4},\cdot), \,\,\,\,\,\,\,\,\,(2|1,1)$ \\
 &&\\
 \hline
 &&\\
 $\braket{\phi_{r -1} \tilde{\mathcal{f}}(\tilde{\lambda}_{k 0},\!\tilde{\phi}_{s 0})}$ & $\tilde{\mathcal{V}}_{ir}^{(1)}$ & $(-\!\frac{3}{2}|\overline{4},\cdot), (-\!1|\overline{4} \!\otimes\! 4,\cdot), (-\!\frac{1}{2}|\overline{4} \!\otimes\!  6, \overline{2}), (0|\overline{4} \!\otimes\!  \overline{4},\cdot), (\frac{1}{2}, \overline{4},\cdot)$ \\
 &&\\
\hline
 &&\\
 $[\tilde{\phi}_{r -1} \tilde{\mathcal{g}}(\tilde{\lambda}_{k 0},\!\tilde{\phi}_{s 0})]$ & $\tilde{\mathcal{V}}_{ir}^{(3)}$ &$(-\!\frac{1}{2}|4,\cdot), \,(0|4 \!\otimes\! 4,\cdot), \,\,(\frac{1}{2}|4 \!\otimes\!  6,2), \,\,(1|4\!\otimes\!  \overline{4},\cdot), \,\,(\frac{3}{2}, 4,\cdot)$ \\
 &&\\
 \hline
 \end{tabular}
 \caption{R Spectrum. The gravitons are assumed to be in an SU(2) singlet and fermions in a 24-dimensional SU(4) representation to be in an SU(2) fundamental representation. All helicity states have double occurrence.}
 \label{RSpectrumTable}
 \end{center} 
  \end{table}
 
 Table \ref{NSSpectrumTable} displays the spectrum in the NS sector. There are no spin $\frac{1}{2}$ or spin 0 particles. Notice that the last two rows have the same physical states as the previous two, with opposite helicities. One pair of these states make up a second graviton, that leads to improper gravitational tree scattering amplitudes, as can be deduced from the fact that in tree scattering amplitudes there are more contractions between matter fields in each helicity set compared to the other graviton, most easily seen in a 3-point MHV amplitude (which vanishes for the second graviton like for conformal gravity). On the other hand, one can argue that the states $\lambda_{i \alpha \, -\!\frac{1}{2}} \tilde{\lambda}_{j \dot{\beta} \, -\!\frac{1}{2}}$, even after gauging the SU(2) symmetry and choosing just one representative, contain two different twistor oscillator modes, one from twistor space and one from the dual, and should lead to Penrose transforms involving these multiple twistors in a symmetric fashion such that there is no guarantee that they refer to massless particles \cite{Penrose:1973, Penrose:1986} (the target space is not limited to ambitwistor space). With other words, $\lambda_{i \alpha \, -\!\frac{1}{2}} \tilde{\lambda}_{j \dot{\beta} \, -\frac{1}{2}}$ might not refer to a massless graviton at all and the plane wave vertex operators $\mathcal{V}_{kj}^{(0)}$ and $\tilde{\mathcal{V}}_{ki}^{(0)}$ might not be suitable. For instance, a more appropriate vertex operator symmetric in the dual twistors might be
 \begin{equation*}
 \mathcal{V} \! \sim \mathcal{G\!F}
   \int \! \frac{\ud t \ud \tilde{t}} {t^2 \tilde{t}^2}\! : \!\prod_{i=1}^2\!\left(\!\delta^2\!(\rho_1 \!-\! t \lambda_i(z)) \mathrm{e}^{it [\tilde{\rho}_1 \mu_i(z)] } \, \delta^2\!(\tilde{\rho}_2 \!-\! \tilde{t} \tilde{\lambda}_i(z)) \mathrm{e}^{i\tilde{t} \braket{\rho_2  \tilde{\mu}_i(z)} } \!\right)\! \!:,
 \end{equation*}
 and we would not have to assign particles with opposite helicities to the same states which by itself is an argument against the appropriateness of the vertex operators $\mathcal{V}_{kj}^{(0)}$ and $\tilde{\mathcal{V}}_{ki}^{(0)}$ in the NS sector (this argument also applies to the use of $\mathcal{V}_{kj}^{(1)}$ and $\tilde{\mathcal{V}}_{ki}^{(1)}$ for the $\phi_{i \alpha \, -\!\frac{1}{2}} \tilde{\phi}_{j  \dot{\beta} \, -\frac{1}{2}}$ states). But so far the author does not have clarity about what these states represent, and table \ref{NSSpectrumTable} stays unaltered for now. In the following we will silently ignore these states for scattering amplitudes.\\
 
 Table \ref{RSpectrumTable} displays the spectrum in the R sector. Here we assume that spin $\frac{1}{2}$ particles belonging to states of the form $[\tilde{\phi}_{i -1} \mathcal{f}(\lambda_{k 0},\!\phi_{s 0})]$, $\!\braket{\phi_{i -1} \mathcal{g}(\lambda_{k 0},\!\phi_{s 0})}, \!\braket{\phi_{i-1} \tilde{\mathcal{f}}(\tilde{\lambda}_{k 0},\!\tilde{\phi}_{s 0})}$, and $[\tilde{\phi}_{i -1} \tilde{\mathcal{g}}(\tilde{\lambda}_{k 0},\!\tilde{\phi}_{s 0})]$ are in the fundamental SU(2) representation. The reason for this is that with this choice the spectrum contains 3 generations of the fermionic particle content of the Pati-Salam model \cite{Pati:1974}\footnote{This assumes that one does not decompose the product $4 \!\otimes\! 6$ or $\overline{4} \!\otimes\! 6$ into irreducible representations $\overline{20} \!\oplus\! \overline{4}$ or $20 \!\oplus\!  4$, respectively.}. It is interesting to note that all helicity states occur twice, reflecting the symmetry between matter fields and their duals. Like in the NS sector vertex operators $\mathcal{V}_{kj}^{(0)}, \tilde{\mathcal{V}}_{kj}^{(0)}, \mathcal{V}_{kj}^{(2)}$ and $\tilde{\mathcal{V}}_{kj}^{(2)}$ with no fermions refer to graviton-like excitations that do not have the proper gravitational tree scattering amplitudes (they have vanishing 3-point MHV amplitudes). Although again these are exactly the vertex operators for states involving more than one twistor oscillator mode usually with no guarantee to be associated with massless particles, the situation is different from the NS sector because the homogeneous functions in the vertex operators are functions of only the zero modes, with the non-zero mode appearing as a single linear factor, such that the interpretation of these states as massless particles of a particular helicity is less controversial.\\
 
 Another interesting point in the R sector is that, when looking at table \ref{RSpectrumTable}, one can notice a strong resemblance with part of the conformal supergravity spectrum of the Berkovits-Witten twistor string with N=4 supersymmetry \cite{Berkovits_1:2004, Dolan:2008}. The main difference is that our model has an additional SU(2) symmetry, it is also symmetric between matter fields and their duals, related with each other through a Fourier transformation \cite{Berkovits_1:2004}, and because of the gauging of the conformal translations it does not contain the states thought of being responsible for the lack of unitarity in the Berkovits-Witten model of conformal supergravity\cite{Berkovits_1:2004}. When considering scattering amplitudes in the next two sections we will silently disregard vertex operators $\mathcal{V}_{ij}^{(2)}, \tilde{\mathcal{V}}_{ij}^{(2)}, \mathcal{V}_{ij}^{(3)},$ and $\tilde{\mathcal{V}}_{ij}^{(3)}$ assuming that they correspond to the same particles as $\tilde{\mathcal{V}}_{ij}^{(0)}, \mathcal{V}_{ij}^{(0)}, \tilde{\mathcal{V}}_{ij}^{(1)},$ and $\mathcal{V}_{ij}^{(1)}$, respectively.\\

\section{EYM Scattering Amplitudes with External Gravitons and Gluons}
  \label{EYM}
  
To compute scattering amplitudes it is advantageous to know simple formulae of integrated vertex operators in addition to fixed ones. Because of the SU(2) symmetry of the bi-twistor all physical states are in some SU(2) representation. To keep it simple, for SU(2) singlet vertex operators, we work in a special gauge and consider just one of the two twistors by disregarding the corresponding $g$ ghosts and omitting the index on the twistor fields. Further down we will present an integrated vertex operator that is BRST invariant under the full BRST charge including the SU(2) ghosts (see \eqref{FermionVertex2}). This will make evident that the gauge-invariant tree and one-loop scattering amplitudes dealt with in this paper are not affected by the gauge choice for the bosonic twistors.\\

In the sector without fermions integrated vertex operators associated to $\mathcal{V}_i$ and $\tilde{\mathcal{V}}_i$ in \eqref{FixedVertex} can be identified with the pure gravitational ones considered in \cite{Kunz:2020}. Using notation \eqref{Notation} they are listed here again, modified to extend them to the full Hilbert space: 
  \begin{equation}
  \begin{split}
  V_i^{(g)} \!=\! \lim_{\eta_{iq} \to 0} \int \!\! \frac{\ud t} {t^2} \ud z & \!\left(\![\tilde{\lambda}(z) \tilde{\rho}_i] - it \, [\tilde{\rho}_i | e(z) \!\ket{\rho_i}
  \frac{\bra{s}\! f(z) | \tilde{\rho}_i]}{\braket{\rho_i s}} \!\right) \\
  &\delta^2\!(\rho_i \!-\! t \lambda(z)) \prod_{q=1}^2 \delta^2(\eta_{iq} - t \phi_q(z)) \,  \mathrm{e}^ {it [\mu(z) \tilde{\rho}_i] } ,\\
  \tilde{V}_i^{(g)} \!=\! \lim_{\tilde{\eta}_{iq} \to 0} \int \!\! \frac{\ud t} {t^2} \ud z & \!\left(\!\braket{\lambda(z) \rho_i } + it \, \!\bra{\rho_i}\! e(z) | \tilde{\rho}_i] 
  \frac{[\tilde{s} | f(z) \!\ket{ \rho_i}}{[\tilde{\rho}_i \tilde{s}]} \!\right)\\
  &\delta^2\!(\tilde{\rho}_i \!-\! t \tilde{\lambda}(z)) \prod_{q=1}^2 \delta^2(\tilde{\eta}_{iq} - t \tilde{\phi}_q(z)) \, \mathrm{e}^{it \braket{\tilde{\mu}(z) \rho_i } } ,\\
  \end{split}
  \label{GRvertex}
  \end{equation}
  where the limit is to be taken at the end when calculating scattering amplitudes, where we used the notation
  \begin{equation}
  \begin{aligned}
  &\!\bra{\rho}\! e(z) |  \tilde{\rho} ] = \rho_{\alpha} e^{\dot{\alpha} \alpha}\!(z) \tilde{\rho}_{\dot{\alpha}}  
  &= \epsilon_{\beta \alpha} \epsilon_{\dot{\alpha} \dot{\beta}} e^{\dot{\alpha} \alpha}\!(z) \rho^{\beta} \tilde{\rho}^{\dot{\beta}} = [ \tilde{\rho} | e(z) \!\ket{\rho} \!,
  \end{aligned}
  \label{Notation}
  \end{equation}
  and where $s^{\alpha}$ and $\tilde{s}^{\dot{\alpha}}$ are reference spinors chosen such that $\braket{\rho_i \, s} \!\ne\! 0 \, (\forall V_{ir})$ and $[\tilde{\rho}_j \, \tilde{s}] \!\ne\! 0 \, (\forall \tilde{V}_{jr})$. The reference spinors serve as projection operators \cite{Kunz:2020}. These extended vertex operators are BRST-closed (when disregarding the SU(2) ghosts and $\vec{H}$ currents of \eqref{Q_current} and after taking the limit) and lead to the same gravitational N$^k$MHV amplitudes  as the ones in \cite{Kunz:2020}, up to a constant.\\
  
  Integrated vertex operators for $\mathcal{V}_{i\cdot}^{(0)}$ and $\tilde{\mathcal{V}}_{i\cdot}^{(0)}$ in \eqref{FixedVertex1} just have an additional factor of $[t \tilde{\lambda}(z) \tilde{\rho}_i]$ or $\braket{t \lambda(z) \rho_i}$ in the integrand:
  \begin{equation}
  \begin{split}
  V_i^{(g0)} \!=\! \lim_{\eta_{iq} \to 0} \int \!\! \frac{\ud t} {t} \ud z &\, [\tilde{\lambda}(z) \tilde{\rho}_i] \!\left(\![\tilde{\lambda}(z) \tilde{\rho}_i] - it \, [\tilde{\rho}_i | e(z) \!\ket{\rho_i}
  \frac{\bra{s}\! f(z) | \tilde{\rho}_i]}{\braket{\rho_i s}} \!\right) \\
  &\delta^2\!(\rho_i \!-\! t \lambda(z)) \prod_{q=1}^2 \delta^2(\eta_{iq} - t \phi_q(z)) \,  \mathrm{e}^ {it [\mu(z) \tilde{\rho}_i] } ,\\
  \tilde{V}_i^{(g0)} \!=\! \lim_{\tilde{\eta}_{iq} \to 0} \int \!\! \frac{\ud t} {t} \ud z  &\, \braket{\lambda(z) \rho_i }\!\left(\!\braket{\lambda(z) \rho_i } + it \, \!\bra{\rho_i}\! e(z) | \tilde{\rho}_i] 
  \frac{[\tilde{s} | f(z) \!\ket{ \rho_i}}{[\tilde{\rho}_i \tilde{s}]} \!\right)\\
  &\delta^2\!(\tilde{\rho}_i \!-\! t \tilde{\lambda}(z)) \prod_{q=1}^2 \delta^2(\tilde{\eta}_{iq} - t \tilde{\phi}_q(z)) \, \mathrm{e}^{it \braket{\tilde{\mu}(z) \rho_i } } ,\\
  \end{split}
  \label{GRvertex1}
  \end{equation}  
    
  Integrated vertex operators associated to $\mathcal{V}_{i}, \tilde{\mathcal{V}}_{i}$ in \eqref{FixedVertex}  and $\mathcal{V}_{ir}^{(1)}, \tilde{\mathcal{V}}_{ir}^{(1)}$  in \eqref{FixedVertex2} describing states with fermions can look like:
  \begin{align}
  V_{ir}^{(x)} \!\!=\!\! \int \!\! \frac{\ud t}{t} \ud z &\! \Biggl[\![\tilde{\lambda}(z\!) \tilde{\rho}_i] \!+\! it \!\left(\![\tilde{\rho}_i | e(z\!) \!\ket{ \rho_i} \!+\!\! \sum_{q=1}^2 [\tilde{\rho}_i | \gamma_{2\cdot q}(z) \!\ket{\eta_{iq}}\!\!\right)\!\!\!
  \left(\!\frac{\bra{s}\! f(z) | \tilde{\rho}_i]}{\braket{\rho_i s}} \!-\!\! \sum_{q=1}^2 \! \frac{\bra{\sigma}\! \beta_{2\cdot q}(z) | \tilde{\rho}_i]}{\braket{\eta_{iq} \sigma}}\right) \!\!\Biggr]\nonumber\\
  &[t \tilde{\phi}_r(z) \tilde{\eta}_{ir}]^x \left([\tilde{\phi}_r(z) \tilde{\eta}_{ir}] + it \sum_{q=1}^2 [\tilde{\eta}_{iq} | \gamma_{1\cdot q}(z) \!\ket{\rho_i}\frac{\bra{s}\! \beta_{1\cdot r}(z) | \tilde{\eta}_{ir}]}{\braket{\rho_i s}}\right) \nonumber\\
  &\delta^2(\rho_i \!-\! t \lambda(z)) \prod_{q=1}^2 \delta^2(\eta_{iq} - t \phi_q(z)) \, \mathrm{exp}\left[it \left([\mu(z) \tilde{\rho}_i] + \sum_{q=1}^2 [\psi_q(z) \tilde{\eta}_{iq}] \right)\right] \,, \label{FermionVertex}\\
  \tilde{V}_{ir}^{(x)} \!\!=\!\! \int \!\! \frac{\ud t}{t} \ud z &\! \Biggl[ \!\braket{\lambda(z\!) \rho_i } \!-\! it \!\left(\!\!\bra{\rho_i}\! e(z) | \tilde{\rho}_i] -\!\!\sum_{q=1}^2 \!\!\bra{\rho_i}\! \gamma_{1\cdot q}(z) | \tilde{\eta}_{iq}]\!\!\right)\!\!\!
  \left(\!\frac{[\tilde{s} | f(z) \!\ket{ \rho_i}}{[\tilde{\rho}_i \tilde{s}]} \!+\!\!\! \sum_{q=1}^2\!\! \frac{[\tilde{\sigma} | \beta_{1\cdot q}(z) \!\ket{\rho_i}}{[\tilde{\eta}_{iq} \tilde{\sigma}]}\right) \!\!\Biggr]\nonumber\\
  &\braket{t \phi_r(z) \eta_{ir}}^x \left(\!\braket{\phi_r(z) \eta_{ir}} + it \sum_{q=1}^2 \!\bra{\eta_{iq}}\! \gamma_{2\cdot q}(z) | \tilde{\rho}_i] \frac{[\tilde{s} | \beta_{2\cdot r}(z) \ket{\eta_{ir}}}{[\tilde{\rho}_i \tilde{s}]}\right) \nonumber\\
  &\delta^2(\tilde{\rho}_i \!-\! t \tilde{\lambda}(z)) \prod_{q=1}^2 \delta^2(\tilde{\eta}_{iq} \!-\! t \tilde{\phi}_q(z)) \, \mathrm{exp}\left[it \left(\braket{\tilde{\mu}(z) \rho_i } +  \sum_{q=1}^2 \!\braket{\tilde{\psi}_q(z) \eta_{iq}}\right)\right] \,,\nonumber
  \end{align}
  where $x \!=\! 0$ when associated to $\mathcal{V}_{i}$ and $\tilde{\mathcal{V}}_{i}$ and $x \!=\! 1$ when associated to $\mathcal{V}_{ir}^{(1)}$ and $\tilde{\mathcal{V}}_{ir}^{(1)}$, and where this time it is additionally assumed that there are Grassmann odd numbers $\epsilon, \epsilon^{\prime}, \tilde{\epsilon}$,  and $\tilde{\epsilon}^{\prime}$ such that $\eta_{iq} = \epsilon \rho_i (\forall V_{ir}, q\!=\!1,\!2)$ and $\tilde{\eta}_{iq} = \tilde{\epsilon} \tilde{\rho}_i (\forall \tilde{V}_{ir}, q\!=\!1,\!2)$, and $\sigma = \epsilon^{\prime} s $ and $\tilde{\sigma} = \tilde{\epsilon}^{\prime} \tilde{s}$ are Grassmann odd reference spinors fulfilling $\frac{\braket{\eta_{iq} \, \sigma}}{\braket{\eta_{jq} \, \sigma}} \!=\! \frac{\braket{\rho_i \, s}}{\braket{\rho_j \, s}} \, (\forall V_{ir}, V_{js}, q\!=\!1,\!2)$ and $\frac{[\tilde{\eta}_{iq} \, \tilde{\sigma}]}{[\tilde{\eta}_{jq} \, \tilde{\sigma}]} \!=\! \frac{[\tilde{\rho}_i \, \tilde{s}]}{[\tilde{\rho}_j \, \tilde{s}]} \, (\forall \tilde{V}_{ir},\tilde{V}_{js}, q\!=\!1,\!2)$. On the other hand, the dual Grassmann spinors should satisfy $\tilde{\eta}_{iq} \ne \tilde{\epsilon} \tilde{\rho}_i (\forall V_{ir}, q\!=\!1,\!2)$ (in addition to $\eta_{iq} = \epsilon \rho_i$) and $\eta_{iq} \ne \epsilon \rho_i (\forall \tilde{V}_{ir}, q\!=\!1,\!2)$ (in addition to $\tilde{\eta}_{iq} = \tilde{\epsilon} \tilde{\rho}_i$) otherwise contractions between fermion fields would vanish in scattering amplitudes. These vertex operators are BRST-closed when disregarding the SU(2) ghosts and $\vec{H}$ currents in \eqref{Q_current}, and contractions from the fermionic ghost and one of the bosonic ghosts between vertex operators of the same helicity set always cancel\footnote{For BRST-closedness alone the Grassmann odd variables $\eta_{iq}$ and $\tilde{\eta}_{iq}$ do not need to be proportional to the $\rho_i$ and $\tilde{\rho}_i$, respectively. For that it is only required that the Grassmann odd spinors $\sigma$ and $\tilde{\sigma}$ are chosen such that $\braket{\eta_{iq} \sigma} \ne 0$ and $[\tilde{\eta}_{iq} \tilde{\sigma}] \ne 0$. The additional conditions are necessary to make the scattering amplitudes manifestly independent of the reference spinors.\label{ReferenceSpinors}}. \\
  
    We take here the opportunity to provide an example of an integrated vertex operator in an SU(2) doublet representation and without gauge fixing the bosonic twistor by rewriting the vertex operator \eqref{FermionVertex} for $x \!=\! 0$ using the SU(2) projection operator:
  \begin{align}
  V_{\rho rt} \!=\!  &\int_{SU(2)} \!\!\!\!\!\!\!\!\!\!\!\ud \mu(g) \frac{\ud t}{t} \ud z \! \Biggl[\![\tilde{\lambda}g^{-1})_r(z\!) \tilde{\rho}_r] \!+\! it \!\left(\sum_{i, j = 1}^2 [\tilde{\rho}_i | (geg^{-1})_{ij}(z) \!\ket{\rho_j} \!+\! \sum_{i, q=1}^2 [\tilde{\rho}_i | (g\gamma_2)_{i q}(z) \!\ket{\eta_{q}}\!\right) \nonumber\\
 &\!\!\left(\sum_{i = 1}^2 \frac{\bra{s_i}\! (gfg^{-1})_{ir}(z\!) | \tilde{\rho}_r]}{2\! \braket{\rho_i s_i}} - \sum_{q = 1}^2 \frac{\bra{\sigma}\! (\beta_2 g^{-1})_{r q}(z\!) | \tilde{\rho}_r]}{\braket{\eta_q \sigma}}\right) \Biggr] \nonumber\\
 &\left([\tilde{\phi}_t(z) \tilde{\eta}_t + it \sum_{i,q=1}^2 [\tilde{\eta}_q | (\gamma_1 g^{-1})_{iq}(z) \!\ket{\rho_i}\frac{\bra{s_i}\! (g\beta_1)_{it}(z) | \tilde{\eta}_t]}{2\! \braket{\rho_i s_i}}\right)  \label{FermionVertex2}\\
 &\prod_{i=1}^2 \delta^2(\rho_i \!-\! t (g\lambda)_i(z)\!) \!\prod_{q=1}^2 \!\delta^2(\eta_{q} \!-\! t \phi_q(z)\!) \mathrm{exp}\Bigl[ it \Bigl( \sum_{i=1}^2 [(\mu g^{-1})_i(z) \tilde{\rho}_i] \!+\!\! \sum_{q=1}^2 [\psi_q(z) \tilde{\eta}_{iq}] \Bigr) \Bigr] ,\nonumber
  \end{align}
  where the indices of $\rho_i, \tilde{\rho}_i, s_i, \tilde{s}_i$ refer to SU(2) vector components, and where $\ud \mu(g)$ is the left- and right-invariant Haar measure of SU(2).
  $V_{\rho rt}$ is indeed BRST invariant under the full BRST charge because of the right-invariance of the Haar measure. \eqref{FermionVertex2} shows that instead of transforming the fields one could equivalently transform the $\rho_i, \tilde{\rho}_j$, and the reference spinors $s_i, \tilde{s}_j$. This way we can get a vertex operator for an SU(2) singlet state by averaging over the $r$ components.\\
    
  Now, we want to consider tree scattering amplitudes in the NS sector with $n$ external gravitons, $k$ of which have negative helicities, and $m$ vector bosons, $q$ of which have negative helicities. Without much justification, more out of convenience and because the scattering amplitudes can be related to EYM amplitudes, we call these vector bosons 'gluons'. We assume SU(2) singlet states for all involved particles and choose vertex operators \eqref{GRvertex} and \eqref{FermionVertex} for gravitons and gluons, respectively. Actually, to make explicit that we only consider vector bosons, we use a simplified version of \eqref{FermionVertex}:
  \begin{align}
  V_{ir} \!\!=\!\! \int \!\! \ud t \ud z &\! \Biggl[\![\tilde{\lambda}(z\!) \tilde{\rho}_i] \!+\! it \!\left(\![\tilde{\rho}_i | e(z\!) \!\ket{ \rho_i} \!+\!\! \sum_{q=1}^2 [\tilde{\rho}_i | \gamma_{2\cdot q}(z) \!\ket{\eta_{iq}}\!\!\right)\!\!\!
  \left(\!\frac{\bra{s}\! f(z) | \tilde{\rho}_i]}{\braket{\rho_i s}} \!-\!\! \sum_{q=1}^2 \! \frac{\bra{\sigma}\! \beta_{2\cdot q}(z) | \tilde{\rho}_i]}{\braket{\eta_{iq} \sigma}}\right) \!\!\Biggr]\nonumber\\
  &\left([\tilde{\phi}_r(z) \tilde{\eta}_{ir}] \sum_{q=1}^2 [\psi_q(z) \tilde{\eta}_{iq}] + \sum_{q=1}^2 [\tilde{\eta}_{iq} | \gamma_{1\cdot q}(z) \!\ket{\rho_i}\frac{\bra{s}\! \beta_{1\cdot r}(z) | \tilde{\eta}_{ir}]}{\braket{\rho_i s}}\right) \nonumber\\
  &\delta^2(\rho_i \!-\! t \lambda(z)) \prod_{q=1}^2 \delta^2(\eta_{iq} - t \phi_q(z)) \, \mathrm{e}^{it [\mu(z)] \tilde{\rho}_i} \,, \label{GluonVertex}
  \end{align}
  and similarly for $\tilde{V}_{ir}$.
          
  Because of bosonic ghost zero modes we need to use fixed vertex operators with regards of the $\gamma_{1,2}$  ghosts. It will turn out that, with at least one external gluon and just using the integrated vertex operators \eqref{GluonVertex} for all gluons (and dividing by an appropriate GL(1)$^2$ factor), the scattering amplitude would not lead to the desired result (it would actually vanish, see at the end of the section), unless we choose the fixed vertex operators representing gluons in a specific way. We use as fixed vertex operators with regards to the $\gamma_{1,2}$ ghosts:
  \begin{align}
  \mathcal{V}_{ir} \!\!= \!\!\!\prod_{l=1 \atop {\alpha, \dot{\alpha}=1}}^2 \!\!\! \delta(\gamma_{1\cdot l}^{\alpha \dot{\alpha}})\!\!
  &\int \!\! \ud t \ud z  \Biggl[\![\tilde{\lambda}(z\!) \tilde{\rho}_i] \!+\! it \!\left(\![\tilde{\rho}_i | e(z\!) \!\ket{ \rho_i} \!+\!\! \sum_{q=1}^2 [\tilde{\rho}_i | \gamma_{2\cdot q}(z) \!\ket{\eta_{iq}}\!\!\right)\nonumber\\
  &\left(\!\frac{\bra{s}\! f(z\!) | \tilde{\rho}_i]}{\braket{\rho_i s}} \!-\!\! \sum_{q=1}^2 \! \frac{\bra{\sigma}\! \beta_{2\cdot q}(z\!) | \tilde{\rho}_i]}{\braket{\eta_{iq} \sigma}}\!\!\right) \!\!\Biggr] 
  \braket{\phi_r(z\!)\chi_{ir}\!} \!\!(\sum_s [\psi_s(z\!) \tilde{\eta}_{is}]) \nonumber\\
  &\delta^2(\rho_i \!-\! t \lambda(z\!)\!) \!\!\prod_{q=1}^2 \!\!\delta^2(\!\eta_{iq} \!-\! t \phi_q(z\!)\!) \mathrm{e}^{it [\mu(z) \tilde{\rho}_i]}  ,\nonumber\\
  \tilde{\mathcal{V}}_{ir} \!\!=\!\!\!\prod_{l=1 \atop {\alpha, \dot{\alpha}=1}}^2 \!\!\! \delta(\gamma_{2\cdot l}^{\dot{\alpha} \alpha})\!\!
  &\int \!\! \ud t \ud z \Biggl[ \!\braket{\lambda(z\!) \rho_i } \!-\! it \!\left(\!\!\bra{\rho_i}\! e(z) | \tilde{\rho}_i] -\!\!\sum_{q=1}^2 \!\!\bra{\rho_i}\! \gamma_{1\cdot q}(z) | \tilde{\eta}_{iq}]\!\!\right) \nonumber\\
  &\left(\!\frac{[\tilde{s} | f(z) \!\ket{ \rho_i}}{[\tilde{\rho}_i \tilde{s}]} \!+\!\!\! \sum_{q=1}^2\!\! \frac{[\tilde{\sigma} | \beta_{1\cdot r}(z) \!\ket{\rho_i}}{[\tilde{\eta}_{ir} \tilde{\sigma}]}\right) \!\!\Biggr]
  \,[\tilde{\phi}_r(z\!) \tilde{\chi}_{ir}] (\sum_s\!\! \braket{\tilde{\psi}_s(z\!) \eta_{is}}\!) \label{FixedGluonVertex}\\
  &\delta^2(\tilde{\rho}_i \!-\! t \tilde{\lambda}(z\!)\!) \!\!\prod_{q=1}^2 \!\!\delta^2(\!\tilde{\eta}_{iq} \!-\! t \tilde{\phi}_q(z\!)\!) \mathrm{e}^{ it \braket{\tilde{\mu}(z) \rho_i } }  ,
  \nonumber
  \end{align}
  where $\chi_{ir}$ and $\tilde{\chi}_{ir}$ are chosen separately from $\eta_{ir}$ and $\tilde{\eta}_{ir}$, otherwise these vertex operators would vanish on the support of the delta functions.\\ 
  
  With these vertex operators a scattering amplitude with at least two external gluons looks like:
  \begin{equation}
  \mathcal{M} = \left< \frac{1}{ \mathrm{vol} \, \mathrm{GL}(2, \mathbb{C})} \, \tilde{\mathcal{V}}_{1r_1} \, \prod_{j=2}^q \! \tilde{V}_{jr_j} \prod_{p=q+1}^{m \!-\! 1} \!\! V_{pr_p} \, \mathcal{V}_{mr_m} \,\, \prod_{i=1}^k \! \tilde{V}_j^{(g)} \prod_{p=k+1}^{n} \!\! V_p^{(g)} \right>_{\!\!\!(g)} \,,
  \label{EYMamplitude}
  \end{equation}
  where the factor $\mathrm{vol} \, \mathrm{GL}(2, \mathbb{C})$ comes from three zero-modes of the $c$ ghost and one zero mode from the $e$ ghosts, and where the subscript $(g)$ at the correlation indicates that one of the $V_p^{(g)}$ and one of the $\tilde{V}_j^{(g)}$ do not participate when taking contractions of twistor fields and fermionic ghost systems.\\
  
  Without any gluons (i.e. with only vertex operators $V^{(g)}$ and $\tilde{V}^{(g)}$) the amplitude reduces to the one considered in \cite{Kunz:2020} yielding the familiar gravitational N$^k$MHV amplitude. Concerning the amplitude \eqref{EYMamplitude} it can be noted that many Wick contractions between vertex operators \eqref{GluonVertex} either get omitted or cancel:
  \begin{itemize}
  \item No contractions occur between integrated vertex operators of different helicity, similarly to the pure gravitational amplitudes \cite{Kunz:2020}, because their contributions cancel with an appropriate choice of reference spinors\footnote{As mentioned in footnote \ref{ReferenceSpinors} the reference spinors can be chosen arbitrarily per vertex operator, but to make the scattering amplitude manifestly independent of them, they must be chosen the same for all vertex operators with the freedom to do it separately for every 'Feynman diagram' contribution to the amplitude. For instance a contraction between an $e$ ghost from a positive helicity vertex $V_{ir}$ with an $f$ antighost from a negative helicity vertex $\tilde{V}_{js}$ contains a factor $[\tilde{\rho}_i \tilde{s}]$ which will be zero by selecting $\tilde{s} = \tilde{\rho}_i$ (still with $[\tilde{s} \tilde{\rho}_j] \ne 0 (\forall \tilde{V}_{jr})$) for this summand of the amplitude.}.
  \item Contractions of $\tilde{\phi}$ and $\psi$ fields between different $V_{ir}$ cancel with contractions between $\gamma_1$ and $\beta_1$ fields and contractions of $\phi$ and $\tilde{\psi}$ fields between different $\tilde{V}_{ir}$ cancel with contractions between $\gamma_2$ and $\beta_2$ fields, such that contractions with the fixed vertex operators \eqref{FixedGluonVertex} are required to make the amplitude non-vanishing.
  \end{itemize}
  It follows that the contractions between fermionic fields are bridged between positive and negative helicities with help of the fixed vertex operators and factorize out as a Parke-Taylor factor with the conventional trace of gauge group generators replaced with products of $\braket{\eta_{ir} \eta_{js}}$ and $[\tilde{\eta}_{kt} \tilde{\eta}_{qu}]$ (compare with (2.7) of \cite{Adamo:2015}). Regarding contractions between twistor fields, they occur between all vertex operators in the same helicity set, but one can argue that graviton exchange between gluons of the same color trace goes beyond tree scattering (a single color trace corresponds to Feynman diagrams in which gluons are all connected and a 'virtual' graviton exchange between them would introduce a loop on a field theoretical level). Although this argument goes beyond the scope of our twistor string model (twistor exchange among gluons are not disallowed in tree scattering), we will skip all twistor and related fermionic ghost correlations between gluons in the same 'color trace'. The result for a 'single trace' (i.e. with no twistor correlations between any gluons) looks then very similar to the EYM amplitude of \cite{Adamo:2015}:
  \begin{align}
  \mathcal{M}_{single \atop{trace}} = & \int \!\!  \prod_{i \in g \cup \tilde{g}} \!\! \frac{\ud t_i} {t_i} \!\!\! \prod_{j \in h \cup \tilde{h}} \!\! \frac{ \ud t_j} {t_j^3}\,\,
  \frac{\mathrm{det}^{\prime} \Phi \,\, \mathrm{det}^{\prime} \tilde{\Phi}}{ \mathrm{vol} \, \mathrm{GL}(2, \mathbb{C})} \mathrm{PT} \prod_{r \in g \cup h} \!\!\! \delta^2(\rho_r \! - \! t_r \lambda(z_r)) \!\! \prod_{l \in \tilde{g} \cup \tilde{h}} \!\!\! \delta^2(\tilde{\rho}_l \! - \! t_l \tilde{\lambda}(z_l)) \label{EYMtree}\\
  & \lim_{\eta_{rs} \to 0 (\forall r \in h) \atop {\tilde{\eta}_{ls} \to 0 (\forall l \in \tilde{h})}} \left< \prod_{r \in g \cup h} \prod_{s=1}^2 \! \delta^2(\eta_{rs} - t_r \phi_s(z_r))  \prod_{l \in \tilde{g} \cup \tilde{h}} \prod_{s=1}^2 \! \delta^2(\tilde{\eta}_{ls} - t_l \tilde{\phi}_s(z_l)) \right> \,, \nonumber
  \end{align}
  where  $g = \{q\!+\!1, \cdots \!, m\}$, $\tilde{g} = \{1, \cdots \!, q\}$,  $h = \{k\!+\!1, \cdots \!, n\}$, $\tilde{h} = \{1, \cdots \!, k\}$, and $\lambda(z_r)$ and $\tilde{\lambda}(z_l)$ fulfill the scattering equations
  \begin{equation*}
  \begin{aligned}
  &\lambda(z_r) = \sum_{l \in \tilde{g} \cup \tilde{h}} t_l \rho_l S(z_r , z_l), &\tilde{\lambda}(z_l) = \sum_{r \in g \cup h} t_r \tilde{\rho}_r S(z_l, z_r),
  \end{aligned}
  \end{equation*}
  where
  \begin{equation}
  S(z_1, z_2) = \frac{\sqrt{\ud z_1}\sqrt{\ud z_2}}{z_1 - z_2}
  \label{Szeg0}
  \end{equation}
  is the Szeg\"o kernel for genus 0.
  $\Phi$ is a symmetric $(n \!-\! k \!+\! 1) \times (n \!-\! k \!+\! 1)$ matrix, and $\tilde{\Phi}$ is a symmetric $(k \!+\! 1) \times (k \!+\! 1)$ matrix
  arising from the Wick contractions between the twistor and ghost fields appearing in the 
  vertex operators with the following elements:
  \begin{equation}
  \begin{aligned}
  &\Phi^{lr} = t_l t_r [\tilde{\rho}_l \tilde{\rho}_r] S(z_l, z_r) \, \text{for} \, l,r \!\in\! h, &\Phi^{l \, |h|\!+\!1} &= \sum_{r \in g} t_l t_r [\tilde{\rho}_l \tilde{\rho}_r] S(z_l, z_r) \, \text{for} \, l \!\in\! h \,,\\
  &\Phi^{ll} \, = - \!\!\! \sum_{r \in h \setminus \{l\}} \!\! \Phi^{lr} \!-\! \Phi^{l \, |h|\!+\!1}  \,\, \text{for} \, l \!\in\! h, &\Phi^{|h|\!+\!1 \, |h|\!+\!1} &= -\!\! \sum_{l \in h} \Phi^{l \, |h|\!+\!1} \,,\\ 
  &\tilde{\Phi}^{lr} = t_l t_r \braket{\rho_l \rho_r} S(z_l, z_r) \, \text{for} \, l,r \!\in\! \tilde{h}, &\tilde{\Phi}^{l \, |\tilde{h}|\!+\!1} &= \sum_{r \in \tilde{g}} t_l t_r \braket{\rho_l \rho_r} S(z_l, z_r) \, \text{for} \, l \!\in\! \tilde{h} \,,\\
  &\tilde{\Phi}^{ll} \, = - \!\!\! \sum_{r \in \tilde{h} \setminus \{l\}} \!\! \tilde{\Phi}^{lr} \!-\! \tilde{\Phi}^{l \, |\tilde{h}|\!+\!1}  \,\, \text{for} \, l \!\in\! \tilde{h}, &\tilde{\Phi}^{|\tilde{h}|\!+\!1 \, |\tilde{h}|\!+\!1} &= -\!\! \sum_{l \in \tilde{h}} \tilde{\Phi}^{l \, |\tilde{h}|\!+\!1} \,.\\
  \end{aligned}
  \label{Matrices}
  \end{equation}
  $\Phi$  and $\tilde{\Phi}$ each have co-rank one with vanishing determinant and det$^{\prime}$ indicates the operation of removing one row and one column before computing the determinant. The result of this operation is actually independent of the choice of row and column removed.\\
  
  Finally, PT in \eqref{EYMtree} denotes a Parke-Taylor factor of the form
  \begin{equation}
  \begin{split}
  \mathrm{PT} &= [\tilde{\chi}_{1r_1} \tilde{\eta}_{q\!+\!1r_1}] S(z_1, z_{q\!+\!1}) \, \braket{\chi_{mr_m} \eta_{qr_m}} \! S(z_m, z_q)\\
  & \prod_{i \in g \setminus \{m\}} \!\! [\tilde{\eta}_{ir_i} \tilde{\eta}_{i\!+\!1r_i}] \!\!\! \prod_{i \in \tilde{g} \setminus \{1\}} \!\!\! \braket{\eta_{i\!-\!1r_{i\!-\!1}} \eta_{ir_{i\!-\!1}}} \!\!\!\! \prod_{i \in g \cup \tilde{g} \setminus \{q,m\}} \!\!\!\!\! S(z_i, z_{i\!+\!1}) \,\,\,\, + \, \text{permutations} \,,
  \end{split}
  \label{PT}
  \end{equation}
  where permutations run over $g$ and $\tilde{g}$ separately per choice of fixed vertex operators.\\
    
  We observe that the $\eta$ and $\tilde{\eta}$ variables in the second row of \eqref{EYMtree} do not occur in the Parke-Taylor factor such that the second row in \eqref{EYMtree} can easily be evaluated to only contribute an overall constant independent of $t_r$ and can be safely disregarded. This way we end up exactly with the formula (2.2) of \cite{Adamo:2015}, up to the numerator of the Parke-Taylor factor. This is a remarkable result.\\
    
  One question to ask is, what would have happened if we had exchanged $\braket{\phi_r(z\!)\chi_{ir}\!}$ $\!\!(\sum_s [\psi_s(z\!) \tilde{\eta}_{is}])$ and $[\tilde{\phi}_r(z\!) \tilde{\chi}_{ir}] (\sum_s\!\! \braket{\tilde{\psi}_s(z\!) \eta_{is}}\!)$ in the fixed vertex operators \eqref{FixedGluonVertex} with $t^{-2}$, as appropriate for a fixed vertex operator belonging to \eqref{GluonVertex}? This would have resulted in full cancellation of contractions between fermions and associated bosonic ghost systems and, therefore, in a vanishing amplitude. This is the reason why we had to introduce the special fixed vertex operators \eqref{FixedGluonVertex}.\\

  In the next section we examine the factorization properties of this EYM amplitude and also of the one-loop amplitudes obtained in \cite{Kunz:2020}.\\

\section{Factorization of Tree and One-Loop Scattering Amplitudes}
  \label{Factorization}
  
  To examine the factorization properties of the scattering amplitudes in section \ref{Spectrum} and in section 3 of \cite{Kunz:2020} we take clues from \cite{Adamo:2015, Adamo:2013}. There are two kinds of worldsheet degenerations: separating and non-separating. The first can be applied to a worldsheet of any genus, but the second only to worldsheets of genus $\ge 1$.
  
\subsection{Separating Degeneration}
\label{Separating}
 We consider first pinching a separating cycle which can be done for both tree and one-loop amplitudes. The degeneration of the Riemann surface $\Sigma$ of genus $g \in \{0,1\}$ (sphere or torus) into a Riemann sphere $\Sigma_L$ and another Riemann surface $\Sigma_R$ of genus g can be modeled in terms of local coordinates by
 \begin{equation*}
 (z_L - z_a)(z_R - z_b) = s \,,
 \end{equation*}
 where $z_a$ is an extra puncture on $\Sigma_L$, $z_b$ is an extra puncture on $\Sigma_R$, and in the pinching limit $s \rightarrow 0$ $z_L$ is located on $\Sigma_L$, $z_R$ on $\Sigma_R$, and the point $z_L = z_a$ on $\Sigma_L$ is glued to the point $z_R = z_b$ on $\Sigma_R$.\\
 
 Near that limit the Szeg\"o kernels behave simply as \cite{Adamo:2015,Adamo:2013}
 \begin{equation*}
 S(z_i, z_j; \tau) = \begin{cases}
                              S(z_i, z_j; \tau) & \text{if} \, z_i, z_j \in \Sigma_L \, \text{or} \, z_i, z_j \in \Sigma_R \,,\\
                              \frac{\sqrt{s}}{\sqrt{\ud z_a} \sqrt{\ud z_b}} S(z_i, z_a) S(z_b, z_j; \tau) + O(s) & \text{if} \, z_i \in \Sigma_L \, \text{and} \, z_j \in \Sigma_R \,,
                             \end{cases}
 \end{equation*}
 where $S(z_i, z_j;\tau)$ stands for any of the torus Szeg\"o kernels when $\Sigma_R$ is a torus and is equal to $S(z_i, z_j)$ in \eqref{Szeg0} when $\Sigma_R$ (and $\Sigma_L$ by setup) is the Riemann sphere.
 
 Similarly, after inserting \cite{Adamo:2015}
 \begin{equation}
 \begin{split}
 &1 = \int \frac{\ud t_a \ud t_b}{\text{vol} \mathbb{C}^{*}} \, \delta \!\left(t_a t_b - \frac{1}{\sqrt{\ud z_a} \sqrt{\ud z_b}}\right) \,,\\
 &1 = \int \ud^2 \lambda_a \, \delta^2(\lambda_a - t_a \lambda(z_a)) \,,\\
 &1 = \int \ud^2 \tilde{\lambda}_b \,\, \delta^2(\tilde{\lambda}_b - t_b \tilde{\lambda}(z_b)) \,, \\
 \end{split}
 \label{Insertions}
 \end{equation}
 the determinants of the matrices \eqref{Matrices} and Parke-Taylor factors \eqref{PT} factorize properly in the degeneration limit. See \cite{Adamo:2015} for details on the EYM tree scattering amplitudes, but it can be easily seen that the argumentation there can trivially be extended to the one-loop amplitudes of \cite{Kunz:2020} which look very similar to the tree scattering amplitude \eqref{EYMtree} without gluons with the Szeg\"o kernels  in gravitational matrices replaced with ones from the torus.
 The result of the factorization is equation (3.2) of \cite{Adamo:2015} with a subset of external momenta going on shell:
 \begin{equation}
 %\mathrm{lim}_{\, \atop{ \, \atop{ \!\!\!\!\!\!\!\!\!\!\!\!\!\!\!\!\!\!\!\!\! (\sum_{\, \atop{\,\atop{\!\!\!\!\!\!\!\!\! i \in L}}} \!\!\!\!\!\!\!\lambda_i \tilde{\lambda}_i)^2 \rightarrow 0}}} \!\!\!\!\!\!
 \!\!\lim_{(\sum_{\,\atop{\,\atop{\!\!\!\!\!\!\!\!\!i \in L}}} \!\!\!\!\!\! \lambda_i \tilde{\lambda}_i)^2 \to 0} \!\!\!\!\!\!\!\!\! 
 \mathcal{M}(\{\lambda_k \tilde{\lambda}_k, \!h_k\}) \!\!=\!\!\! \sum_{h = \pm} \! \int \!\! \frac{\ud^2 \lambda \ud^2 \tilde{\lambda}}{\mathrm{vol} \mathbb{C}^{*}} \mathcal{M}_L(\{\lambda_i \tilde{\lambda}_i, h_i\}_{i \in L} ; \!\lambda \tilde{\lambda}, \!h) \mathcal{M}_R(\!-\!\lambda \tilde{\lambda}, \!-\!h; \!\{\lambda_j \tilde{\lambda}_j, \!h_j\}_{j \in R}).
 \label{FactorizedAmplitude}
 \end{equation}
 
 Applying this to the one-loop scattering amplitude of \cite{Kunz:2020} in the NS sector, $\mathcal{M}$ and $\mathcal{M}_R$ are of the form
 \begin{equation}
 \begin{split}
 \mathcal{M}(\{\lambda_j \tilde{\lambda}_j, \!h_j\}) \!\!&=\!\! \int \!\! \frac{\ud \tau}{\mathrm{Im}\tau \, \mathrm{vol} \mathbb{C}^{*}} \!\prod_{j = 1}^n  \frac{\ud t_j} {t_j^3}\,\,
   \!\!\prod_{k \in \{+\}} \! \delta^2(\lambda_k \! - \! t_k \lambda(z_k)) \prod_{l \in \{-\}} \!  \delta^2(\tilde{\lambda}_l \! - \! t_l \tilde{\lambda}(z_l))\\
 &\sum_{\alpha = 2,3,4} \!\! \mathrm{det} \Phi_{\alpha} \mathrm{det} \tilde{\Phi}_{\alpha}\,\, \mathcal{Z}(\tau) \,,
 \end{split}
 \label{Oneloop}
 \end{equation}
 where 
 \begin{equation*}
  \begin{split}
  &\tilde{\Phi}_{\alpha}^{lm} = t_l t_m \braket{\lambda_l \lambda_m } S_1(z_l, z_m; \tau), \, l \ne m,\\
  &\tilde{\Phi}_{\alpha}^{\,ll} = - \sum_{m \neq l} t_l t_m \braket{\lambda_l \lambda_m } S_{\alpha}(z_l, z_m; \tau)
  \end{split}
  \end{equation*}
  for $l,m \in \{-\}$ = the subset of negative helicities, and
  \begin{equation*}
  \begin{split}
  &\Phi_{\alpha}^{rs} = t_r t_s \, [\tilde{\lambda}_r \tilde{\lambda}_s] \, S_1(z_r, z_s; \tau), \, r \ne s,\\
  &\Phi_{\alpha}^{rr} = - \sum_{s \neq r} t_r t_s [\tilde{\lambda}_r \tilde{\lambda}_s] S_{\alpha}(z_r, z_s; \tau)
  \end{split}
  \end{equation*}
  for $r,s \in \{+\}$ = the subset of positive helicities, $S_1(z_i, z_j; \tau)/S_{\alpha}(z_i, z_j; \tau)$ are the torus Szeg\"o kernels for odd/even spin structure, $\mathcal{Z}(\tau) = \mathrm{Im}\tau^{-6}( \eta(\tau) )^{-24}$ is the modular-invariant contribution from the one-loop partition function with $\eta(\tau)$ being the Dedekind eta function, the $\tau$ integration can be limited to the fundamental region because of modular invariance, and $\mathcal{M}_R(\cdots ,\!-\!h; \cdots)$ is the same as $\mathcal{M}$ with an additional external puncture in $\{\pm\}$ for $h \lessgtr 0$.
  The tree amplitude $\mathcal{M}_L$ is given by (compare with \eqref{EYMtree} for $g \cup \tilde{g} = \emptyset$)
  \begin{equation*}
  \mathcal{M}_{tree} = \int \! \prod_{j =1}^n \! \frac{ \ud t_j} {t_j^3}\,\,
  \frac{\mathrm{det}^{\prime} \Phi_0 \,\, \mathrm{det}^{\prime} \tilde{\Phi}_0}{ \mathrm{vol} \, \mathrm{GL}(2, \mathbb{C})}  \prod_{r \in \{+\}} \!\!\! \delta^2(\lambda_r \! - \! t_r \lambda(z_r)) \!\! \prod_{l \in \{-\}} \!\!\! \delta^2(\tilde{\lambda}_l \! - \! t_l \tilde{\lambda}(z_l)) \,,
  \end{equation*}
  where $\Phi_0$ is like $\Phi_1$, but with all occurrences of $S_1(z_i, z_j; \tau)$ replaced with $S(z_i, z_j)$, likewise for $\tilde{\Phi}_0$, and where $\mathcal{M}_L(\cdots ,h)$ is the same as $\mathcal{M}_{tree}$ with an additional external puncture in $\{\pm\}$ for $h \gtrless 0$. This shows that the one-loop scattering amplitude has the correct separating factorization properties which are essential for unitarity.\\
 
 The analogue situation applies to the one-loop scattering amplitude for odd torus spin structures in the R sector \cite{Kunz:2020}, but it needs to be pointed out that the spin structures on the sphere and the torus are different, i.e. we are not dealing with NS punctures, but R punctures, such that the fields satisfy scattering equations on the cylinder and not on the Riemann sphere, and that the gluing procedure is 'twisted' between the Riemann sphere and the torus. More about this can be found in the second half of the next subsection.\\
 
Unitary factorization requires to consider complete sets of intermediate states, but for the scattering amplitudes considered here there was no need to include additional intermediate states because the corresponding tree scattering amplitudes vanish due to dangling matter fields that cannot get contracted.\\

\subsection{Non-separating Degeneration}
\label{NonSeparating}
 We want to check in more detail whether the one-loop amplitude \eqref{Oneloop} factorizes properly for a non-separating degeneration of the worldsheet. Pinching a non-separating cycle occurs in the limit $q = e^{2\pi i \tau} \rightarrow 0$. One obstruction to dealing with this limit by simply taking the residue at 0 of the integrand as a function in $q$ is that the integrand is not an analytic function in $q$ because it has an $\mathrm{Im}(\tau)$ in the denominator and the torus Szeg\"o kernel $S_1(z_i, z_j; \tau)$ also contains such a term:
  \begin{equation}
  S_1(z_i, z_j; \tau) = \left(\frac{\theta_1^{'}\!(z_i - z_j; \tau)}{\theta_1(z \! - \! z_i; \tau)} + 4 \pi \frac{\mathrm{Im}(z_i \! - \! z_j)}{\mathrm{Im}(\tau)} \right) \sqrt{\ud z_i}\sqrt{\ud z_j} \,.
  \end{equation} 
  The term in the denominator of the integrand of \eqref{Oneloop} is actually not entirely correct because it arises from disregarding the zero mode of the $c$ ghost replacing it with a $\ud z$ integration and discarding a $\ud k$ momentum integration in the partition function \cite{Kunz:2020}, i.e. $\mathrm{Im}\tau^{-1}$ should be replaced with $\ud k \ud z^{-1}$, what still keeps the modular invariance of the amplitude intact. The behavior of $S_1$ can be corrected by modeling the torus as a Riemann sphere with a handle with ends located at two reference points $\sigma_a$ and $\sigma_b$, i.e. we add these two points as additional punctures, $\sigma_a$ to the set with positive helicities and $\sigma_b$ to the set of negative helicities, modeled like in the separating case in terms of local coordinates by
 \begin{equation}
 (\sigma - \sigma_a)(\sigma - \sigma_b) = q.
 \label{factorization}
 \end{equation}
 Then, because according to \cite{Tuite:2010} all Szeg\"o kernels are holomorphic in $q$ for small enough $q$, the integrand in \eqref{Oneloop} becomes a manifestly meromorphic function in $q$, and we can take the residue in the pinching limit $q \rightarrow 0$.\\
  The asymptotic behavior of the Szeg\"o kernels is \cite{Casali:2014, Geyer:2016}:  
 \begin{equation}
 \begin{aligned}
 &S_{1,2}(z_i, z_j; \tau) \rightarrow &\frac{1}{2}S(\sigma_i, \sigma_j)&\left(\sqrt{\frac{S(\sigma_i, \sigma_a)S(\sigma_j, \sigma_b)}{S(\sigma_i, \sigma_b)S(\sigma_j, \sigma_a)}} + \sqrt{\frac{S(\sigma_i, \sigma_b)S(\sigma_j, \sigma_a)}{S(\sigma_i, \sigma_a)S(\sigma_j, \sigma_b)}} \right) &+\, O(q),\\
 &S_{3,4}(z_i, z_j; \tau) \rightarrow &S(\sigma_i, \sigma_j)  \pm &\sqrt{q} \frac{S(\sigma_i, \sigma_a)S(\sigma_j, \sigma_b)S(\sigma_i, \sigma_b)S(\sigma_j, \sigma_a)} {S(\sigma_i, \sigma_j)S(\sigma_a, \sigma_b)^2}  &+\, O(q)\,\,\\
 &&= S(\sigma_i, \sigma_j)  &\pm \sqrt{q} \frac{S(\sigma_i, \sigma_a)S(\sigma_j, \sigma_b) - S(\sigma_i, \sigma_b)S(\sigma_j, \sigma_a)}{S(\sigma_b, \sigma_a)}  &+\, O(q),
 \end{aligned}
 \label{asymptotic}
 \end{equation} 
 in terms of the coordinates $\sigma = e^{2 \pi i (z \!-\! \frac{\tau}{2})}$ on the Riemann sphere and where $S(\sigma_i, \sigma_j)$ is the Szeg\"o kernel \eqref{Szeg0} for genus 0 \footnote{The limiting behavior of the torus Szeg\"o kernels given here is more appropriate than the one stated in \cite{Kunz:2020}.}. For $q \rightarrow 0$ the partition function behaves as $\sim q^{-1}$ and $\ud \tau \sim \ud q/q$ such that the leading behavior of the integrand in \eqref{Oneloop} is controlled by the term
 \begin{equation*}
 \frac{\ud q}{q^2}  \sum_{\alpha = 2,3,4}  \mathrm{det} \Phi_{\alpha} \mathrm{det} \tilde{\Phi}_{\alpha} \,.
 \end{equation*}
  
  Because the Szeg\"o kernels $S_1$ and $S_2$ have the same limit, the matrices $\Phi_2$ and $\tilde{\Phi}_2$ degenerate into matrices representing tree scattering of the $n$ particles with vanishing determinants, such that $\mathrm{det} \Phi_2 \mathrm{det} \tilde{\Phi}_2 \sim q^2$ and the leading term for $\alpha = 2$ does not contribute at all to the amplitude. Regarding the matrices $\Phi_{3,4}$ and $\tilde{\Phi}_{3,4}$, the difference between the $O(1)$ limit of the Szeg\"o kernel $S_1$ and $S_{3,4}$ seems to indicate that the matrices do not degenerate in the $q \rightarrow 0$ limit, but one can argue that in the leading term the dependence on $\sigma_a$ and $\sigma_b$ is located purely in the asymptotic behavior \eqref{asymptotic} of $S_1$ and the term cannot depend on the actual value of $\sigma_a$ and $\sigma_b$, and as the punctures $\sigma = e^{2 \pi i (z \!-\! \frac{\tau}{2})}$ move away in the limit, we can take the asymptotic behavior of $S_1$ as independent of $\sigma_a$ and $\sigma_b$ (or in the limit $\sigma_{a,b} \rightarrow \infty$) up to $O(q)$:
  \begin{equation}
  S_1(\sigma_i, \sigma_j; \tau) \rightarrow S(\sigma_i, \sigma_j) + O(q)\\
  \label{S_1}
  \end{equation} 
  such that the leading term of $S_1$ coincides with the one of $S_{3,4}$ and the matrices degenerate again. But this time $\mathrm{det} \Phi_{3,4} \mathrm{det} \tilde{\Phi}_{3,4} \sim q$ and the terms for $\alpha = 3,4$ provide non-vanishing contributions by replacing a propagator between two particles with a pair of propagators connecting these particles with one of the reference points $\sigma_{a,b}$. That leads to the interpretation that there are two additional particles that are being scattered, one flowing from $\sigma_a$ to $\sigma_b$ and the other one flowing from $\sigma_b$ to $\sigma_a$ with opposite momenta. We do the same insertions \eqref{Insertions} into \eqref{Oneloop} as in the separating case and also 'trade' a $\delta(1 \!-\! (t_a t_b \sqrt{\ud \sigma_a} \sqrt{\ud \sigma_b})^{-1})$ for an additional (vol$\mathbb{C}^{*})^{-1}$ \cite{Adamo:2015}, such that \eqref{Oneloop}  becomes (also changing (vol$\mathbb{C}^{*})^3$ in the denominator to (vol$\mathbb{C}^{*}$ GL(2,$\mathbb{C}$)) because of ghost zero modes for a genus 0 worldsheet, and also reverting to use $z$ instead of $\sigma$ for Riemann sphere coordinates):
 \begin{equation}
 \begin{split}
 \mathcal{M} \!=\!\! \int \!\! \frac{\ud^2 \!\lambda_a  \ud^2 \!\tilde{\lambda}_b}{\mathrm{vol} \mathbb{C}^{*}}&\int \!\! \frac{1}{\mathrm{vol GL}(2, \mathbb{C})} \!\! \prod_{j  \in \{+,a\} \atop{\cup \{-,b\}}} \!\!\frac{\ud t_j} {t_j^3} \!\!
 \prod_{k \in \{+,a\}} \!\!\! \delta^2\!(\lambda_k \! - \! t_k \lambda(z_k)) \!\!\! \prod_{l \in \{-,b\}} \!\!\! \delta^2\!(\tilde{\lambda}_l \! - \! t_l \tilde{\lambda}(z_l))\\
 &\!\!\sum_{\alpha = 3,4} \lim_{q \to 0} \frac{1}{q} \left(t_a t_b\mathrm{det} \Phi_{\alpha} \, t_a t_b \mathrm{det} \tilde{\Phi}_{\alpha}\right)\,.
 \end{split}
 \label{LimitNSoneloop}
 \end{equation}
 The last term can be represented more tellingly by noticing that the $\sqrt{q}$ term of  $S_{3,4}$ in \eqref{asymptotic} multiplies a subdeterminant of a matrix that arises from deleting a row and a column of a tree scattering matrix, i.e. it is equal to the det$^{\prime}$ operation  in \eqref{EYMtree}. Therefore,
 \begin{equation*}
 \begin{split}
 &\lim_{q \to 0} \frac{t_a t_b}{\sqrt{q}} \mathrm{det} \Phi_{\alpha} \sim \sum_{r,s \in \{+\} \atop{r \ne s}} t_a t_b  t_r t_s \,[\tilde{\lambda}_r \tilde{\lambda}_s]\,\, \frac{S(z_r, z_a)S(z_s, z_b) - S(z_r, z_b)S(z_s, z_a)} {S(z_b, z_a)}  \mathrm{det}^{\prime} \Phi_{tree} \,,\\
 &\lim_{q \to 0} \frac{t_a t_b}{\sqrt{q}} \mathrm{det} \tilde{\Phi}_{\alpha}  \sim \sum_{l,m \in \{-\} \atop{l \ne m}} \! t_a t_b t_l t_m \!\braket{\lambda_l \lambda_m}\! \frac{S(z_l, z_b)S(z_m, z_a)- S(z_l, z_a)S(z_m, z_b)} {S(z_a, z_b)} \mathrm{det}^{\prime} \tilde{\Phi}_{tree} \,,
 \end{split}
 \end{equation*}
 where $\Phi_{tree}$ and $\tilde{\Phi}_{tree}$ arise from $\Phi_{\alpha}$ and $\tilde{\Phi}_{\alpha}$ by replacing all propagators with a propagator $S$ for genus 0, respectively. Under the support of the scattering equation for $\lambda_a$ and $\tilde{\lambda}_b$ this reduces to
 \begin{equation}
 \begin{split}
 &\lim_{q \to 0} \frac{t_a t_b}{\sqrt{q}} \mathrm{det} \Phi_{\alpha} \sim \frac{2 \sqrt{\ud z_b}}{S(z_b, z_a)} \sum_{r \in \{+\}} t_a t_r \,[\tilde{\lambda}_b \tilde{\lambda}_r]\,\,S(z_r, z_a) \, \mathrm{det}^{\prime} \Phi_{tree} \,, \\
 &\lim_{q \to 0} \frac{t_a t_b}{\sqrt{q}} \mathrm{det} \tilde{\Phi}_{\alpha}  \sim \frac{2 \sqrt{\ud z_a}}{S(z_a, z_b)} \sum_{l \in \{-\}} t_b t_l \, \braket{\lambda_a \lambda_l} \,\, S(z_l, z_b) \, \mathrm{det}^{\prime} \tilde{\Phi}_{tree} \,.
 \end{split}
 \label{NonSepFactorization}
 \end{equation}
 By introducing additional delta functions through
 \begin{equation*}
 1 = \int \ud^2 \!\tilde{\lambda}_a  \ud^2 \!\lambda_b \,\, \delta^2(\lambda_b \!+\! \lambda_a)\, \delta^2(\tilde{\lambda}_a \!-\! \tilde{\lambda}_b)
 \end{equation*}
 the interpretation of the particle exchange with opposite momenta becomes manifest.\\
 
 This shows explicitly that the one-loop scattering amplitude \eqref{Oneloop} for even spin structures has good factorization properties in the non-separating degeneration limit. But an even more convincing result (required for establishing unitarity) would be if the limit coincided with a single unitary cut of a tree amplitude with $n+2$ particles, like
 \begin{equation}
 \mathcal{M}_{k,n}^{1-loop} = \! \int \!\! \frac{\ud^2 \lambda_0 \ud^2 \tilde{\lambda}_0}{\mathrm{vol} \mathbb{C}^{*}}\int \frac{\ud q}{q^2} \mathcal{M}_{k+1,n+2}^{tree}(q; \{\lambda_i, \tilde{\lambda}_i ; \lambda_0, \tilde{\lambda}_0 \}_{i \in \{-\}} ; \! \{\lambda_j, \tilde{\lambda}_j ; -\!\lambda_0, \tilde{\lambda}_0 \}_{j \in \{+\}}) \,,
 \label{unitary_cut}
 \end{equation}
 where the subscript $k,\!n$ means $k$ particles with negative helicity and $n-k$ particles with positive helicity, and the $q$-dependence of $\mathcal{M}^{tree}$ refers to the modeling equation \eqref{factorization} with the interpretation that in the limit $q \rightarrow 0$ punctures become infinitely separated from the puncture of the loop particle belonging to the other helicity set ('long handle'). Expressing $\mathcal{M}^{tree}$ in terms of the homogenous coordinates $\sigma = \frac{1}{t}(1,z)$ (not to be confused with the earlier use of $\sigma$ for Riemann sphere coordinates) the degeneration can be expressed by scaling the second coordinate $\sigma_a^2 = \frac{z_a}{t_a} \sim q^{-\frac{1}{2}} \sigma_a^2$ and $\sigma_b^2 \sim q^{-\frac{1}{2}} \sigma_b^2$ when referring to the (remote) loop location in scattering equations but not in propagators in the matrices due to contractions between particles of the same helicity set.
 The scattering equations for $\lambda_0$ and $\tilde{\lambda}_0$ then become
 \begin{equation*}
 \begin{split}
 \lambda_0 \!=\!\!\! \sum_{i \in \{-\}} \frac{\lambda_i }{(i a)} \!-\! \frac{\lambda_0}{(b a)} = \frac{\sqrt{q}}{\sigma_a^2} \sum_{i \in \{-\}} \frac{\lambda_i}{\sigma_i^1} + O(q) \,, \\
 \tilde{\lambda}_0 \!=\!\!\! \sum_{i \in \{+\}}\frac{\tilde{\lambda}_i}{(i b)} \!+\! \frac{\tilde{\lambda}_0}{(a b)} = \frac{\sqrt{q}}{\sigma_b^2} \sum_{i \in \{+\}} \frac{\tilde{\lambda}_i}{\sigma_i^1} + O(q) \,,
 \end{split}
 \end{equation*}
 where we used the notation $(ij) = \sigma_i \cdot \sigma_j = \sigma_i^1 \sigma_j^2 - \sigma_i^2 \sigma_j^1$.
 
 From this it follows that the following elements in the matrices scale as $\sqrt{q}$:
 \begin{equation*}
 \begin{split}
 &\Phi^{a j} = \frac{[0 \, j]} {(a \, j)} = \sqrt{q} \frac{1}{\sigma_b^2} \sum_{i \in \{+\}} \frac{[i \, j]} {\sigma_i^1 (a \, j)} + O(q), \,\, j \in \{+\} \,, \\
 &\tilde{\Phi}^{b \, l} = \frac{\braket{0 \, l}} {(b \, l)} = \sqrt{q} \frac{1}{\sigma_a^2} \sum_{m \in \{-\}} \frac{\braket{m l}}{\sigma_m^1 (b \, l)} + O(q), \,\,\,  l \in \{-\} \,,
 \end{split}
 \end{equation*} 
 with the notation $[i \, j] = [ \tilde{\lambda}_i \tilde{\lambda}_j]$, $\braket{i \, j} = \braket{\lambda_i \lambda_j}$. The amplitude then becomes
 \begin{equation}
 \begin{split}
 \mathcal{M}_{k,n}^{1-loop} \!=\!\! \int \!\! \frac{\ud^2 \!\lambda_0  \ud^2 \!\tilde{\lambda}_0}{\mathrm{vol} \mathbb{C}^{*}}&\int \!\! \frac{1}{\mathrm{vol GL}(2, \mathbb{C})} \!\! \prod_{j  \in \{+,a\} \atop{\cup \{-,b\}}} \!\! \ud^2 \sigma_j \!\!
 \prod_{s \in \{+,a\}} \!\!\! \delta^2\!(\lambda_s \! - \! \lambda(\sigma_s)) \!\!\! \prod_{l \in \{-,b\}} \!\!\! \delta^2\!(\tilde{\lambda}_l \! - \! \tilde{\lambda}(\sigma_l))\\
 &\lim_{q \to 0} \frac{1}{q} \left(\mathrm{det}^{\prime} \Phi_{n-k+1}^{tree} \, \mathrm{det}^{\prime} \tilde{\Phi}_{k+1}^{tree}\right)
 \end{split}
 \label{single_cut}
 \end{equation}
 with
 \begin{equation}
 \begin{split}
 &\lim_{q \to 0} \frac{1}{\sqrt{q}} \mathrm{det}^{\prime} \Phi_{n-k+1}^{tree} \!\! = \frac{1}{\sigma_b^2} \! \sum_{i, j\in \{+\} \atop {i \ne j}} \! \frac{[j \, i]} {\sigma_j^1 (a \, i)} \mathrm{det}^{\prime} \Phi_{n-k}^{tree} =  \!\! \sum_{i \in \{+\} } \lim_{(j b) \to  \sigma_j^1 \sigma_b^2 \atop{(\forall j \in \{+\})}} \frac{[0 \, i]} {(a \, i)} \mathrm{det}^{\prime} \Phi_{n-k}^{tree}  \,, \\
 &\lim_{q \to 0} \frac{1}{\sqrt{q}} \mathrm{det}^{\prime} \tilde{\Phi}_{k+1}^{tree} = \frac{1}{\sigma_a^2} \sum_{l,m \in \{-\} \atop {l \ne m}} \frac{\braket{l \, m}} {\sigma_l^1 (b \, m)} \mathrm{det}^{\prime} \tilde{\Phi}_{k}^{tree} = \!\!\!\sum_{m \in \{-\}}  \lim_{(l a) \to  \sigma_l^1 \sigma_a^2 \atop{(\forall l \in \{-\})}} \frac{\braket{0 \, m}} {(b \, m)} \mathrm{det}^{\prime} {\tilde\Phi}_k^{tree} \,,
 \end{split}
 \label{single_cut_matrices}
 \end{equation}
 where $\Phi_{n-k}^{tree}$ and $\tilde{\Phi}_{k}^{tree}$ are the n-particle tree amplitude matrices obtained from $\Phi_{n-k+1}^{tree}$ and $\tilde{\Phi}_{k+1}^{tree}$ by deleting all entries referring to $\sigma_a$ and $\sigma_b$ and deleting the row and column containing only zero entries, respectively.
 
 Now, the equations in \eqref{NonSepFactorization}, when expressed in homogeneous coordinates, are exactly the same as in \eqref{single_cut_matrices}, up to scaling factors proportional to $z_a \!-\! z_b$ (which can be viewed as arising from ghost zero modes at the loop particle punctures absorbed into vol GL$(2, \mathbb{C})$) and taking the limit.
 
 However, to establish unitary factorization, we must not just consider intermediate gravitons, but rather a complete set of physical states. Intermediate states from the R sector do not contribute, because a tree scattering amplitude with just two R states, one with positive and the other one with negative helicity, and any number of NS states vanishes due to vanishing determinants of the resulting matrices of co-rank 1 \footnote{Tree scattering amplitudes containing at least three external states from both NS and R sector are more complicated because the vertex operators in both sectors share the same ghosts. Such amplitudes are not considered here.}. In order to extend the unitary cut for loop particles from gravitons to other states in the NS sector, namely gluons and fermions, it can be noticed that in \eqref{Matrices} the matrices for a single gluon or spin $\frac{3}{2}$ fermion look the same as for an additional graviton such that \eqref{single_cut} just would pick up as another factor the Parke-Taylor PT factor multiplied with $t_a^2 t_b^2$, which is $(a b)^{-2}$ for a single gluon or fermion. The same factor would appear in \eqref{LimitNSoneloop} because for these loop particles one would have to insert the corresponding on-shell fixed vertex operators at $z_a$ and $z_b$ \cite{Adamo_1:2013}\footnote{This was also true for the pure gravitational loop particles, but because of the simple form of the vertex operators \eqref{FixedVertex} they were fully taken care of by adjusting the scattering equations for two additional particles at $z_a$ and $z_b$ (compare with \eqref{OnShellVertices}).}. This shows that we have indeed unitary factorization \eqref{unitary_cut} for a complete set of intermediate states.\\
 
 A similar procedure can be done on the torus with odd spin structure for a non-separating factorization of the amplitude \cite{Kunz:2020} (notice that, as mentioned in section \ref{Spectrum}, this amplitude does not represent a proper gravitational amplitude):
 \begin{equation}
  \begin{split}
  \mathcal{M} = \, & \delta^4(\sum_{i=1}^n \lambda_i \tilde{\lambda}_i) \int \! \frac{\ud\tau \, \ud^2\lambda_0 \ud^2\tilde{\lambda}_0}{\text{vol} \mathbb{C}^{*}} \prod_{i=1}^n  
  \frac{\ud t_i \ud z_i} {t_i^3}\,\, \prod_{l=1}^k \!\ \delta^2(\tilde{\lambda}_l \! - \! t_l \tilde{\lambda}(z_l)) \!\!\prod_{r=k+1}^n \!\!\! \delta^2(\lambda_r \! - \! t_r \lambda(z_r)) \\
  &\prod_{r \in \{+\}} \!\!\Phi_{rr} \, \mathrm{det} \Phi_1 \prod_{l \in \{-\}}\!\! \tilde{\Phi}_{ll} \, \mathrm{det} \tilde{\Phi}_1 \,\,\mathcal{Z}(\tau) \,,
  \end{split}
  \label{Ramplitude}
  \end{equation}
  where for $l,m \in \{-\}$, and $r,s \in \{+\}$:
  \begin{align*}
  &\tilde{\Phi}_1^{lm} = \tilde{\Phi}_{\alpha}^{lm} = t_l t_m \braket{l m} \frac{S_1(z_l, z_m; \tau)}{\sqrt{\ud z_l \ud z_m}}, \, l \ne m \,,\\
  &\tilde{\Phi}_1^{\,ll} = - t_l \braket{l 0} - \sum_{m \neq l} \tilde{\Phi}_1^{lm} = -t_l \, \mathrm{lim}_{z \rightarrow z_l} \!\! \braket{l \lambda(z)},\\
  &\Phi_1^{rs} = \Phi_{\alpha}^{rs} = t_r t_s [r s] \frac{S_1(z_r, z_s; \tau)}{\sqrt{\ud z_r \ud z_s}}, \, r \ne s \,,\\
  &\Phi_1^{rr} = - t_r [r 0] - \sum_{s \neq r} \Phi_1^{rs} = -t_r \, \mathrm{lim}_{z \rightarrow z_r} [ r \tilde{\lambda}(z)] \,,
  \end{align*}
  with notation similar to the NS one-loop amplitude \eqref{Oneloop}\footnote{In \cite{Kunz:2020} the products of the diagonal matrix elements were missing because the wrong  vertex operators \eqref{GRvertex} (appropriate for the NS sector) instead of \eqref{GRvertex1} (appropriate for the R sector) were used although this does not affect the factorization properties critically. Notice that compared to the NS sector here the zero mode fields $\ud^2 \lambda_0$ and $\ud^2 \tilde{\lambda}_0$ replace the factor Im$\tau^{\!-\!1} \sim \ud k \ud z^{\!-\!1}$.}.
 This time the torus can be viewed as arising from sewing a cylinder to itself or a Riemann sphere to itself in twisted fashion, using the modeling equation \eqref{factorization} expressed on the cylinder:
 \begin{equation}
 \frac{\sqrt{\ud z \ud z_a}}{S_{cyl}(z - z_a)} \frac{\sqrt{\ud z \ud z_b}}{S_{cyl}(z - z_b)} = q \,.
 \label{Rfactorization}
 \end{equation}
 We want to show that the amplitude \eqref{Ramplitude} can be obtained from the single unitary cut of an $n+2$ particle tree amplitude on the cylinder.\\
  
  In the $q \rightarrow 0$ limit the Szeg\"o kernel $S_1$ behaves as in \eqref{asymptotic} but, again, the dependence on $\sigma_a$ and $\sigma_b$ can be disregarded in the leading term, i.e. we use equation \eqref{S_1} again, taking into account that $S(\sigma_i, \sigma_j) = S_{cyl}(z_i, z_j)$ for $\sigma = e^{2\pi iz}$. The determinants of the matrices $\Phi_1$ and $\tilde{\Phi}_1$ vanish up to terms that multiply the zero modes $\tilde{\lambda}_0$ and $\lambda_0$, respectively.  Again, we make the interpretation that there are two additional particles scattered between $\sigma_a$ and $\sigma_b$ with opposite momenta, and we insert two on-shell fixed vertex operators \eqref{FixedVertex1}, one at $z_a$ and one at $z_b$, leading to the insertion:
  \begin{equation}
  1 = \int \frac{\ud t_a \ud t_b}{\text{vol} \mathbb{C}^{*}} \, \delta \!\left(t_a t_b - \frac{1}{\sqrt{\ud z_a} \sqrt{\ud z_b}}\right) \left[t_a [\tilde{\lambda}_a \tilde{\lambda}(z_a)] \, t_b \!\braket{\lambda_b \lambda(z_b)} + \cdots \right] ,
  \label{OnShellVertices}
  \end{equation}
  where $\tilde{\lambda}_a$ and $\lambda_b$ are the same as $\tilde{\lambda}_0$ and $\lambda_0$ up to a factor soon to be determined and where $\cdots$ stands for additional factors from on-shell vertex operators making up a complete set of physical states in the R sector \footnote{The NS sector does not contribute because, like before but with NS and R exchanged, a tree scattering amplitude with just two NS states, one from each helicity set, and any number of R states vanishes due to vanishing determinants of the resulting matrices of co-rank 1}. We skip the additional states for now and mention them again at the end.
    The scattering equations can then be changed to  
  \begin{align}
  \lambda_j =& \, t_j \lambda_0 + \sum_{i \in \{-\}} t_j t_i \lambda_i \frac{S_1(z_j, z_i; \tau)}{\sqrt{\ud z_j} \sqrt{\ud z_i}} &\rightarrow  && t_j \lambda_0  + \sum_{i \in \{-\}} \frac{t_j t_i \lambda_i \sqrt{\sigma_j\sigma_i}}{\sigma_j - \sigma_i} + O(q) =  \nonumber\\
  &\frac{t_j t_b \lambda_b}{- \sqrt{\sigma_b}}  + \sum_{i \in \{-\}} \frac{t_j t_i \lambda_i \sqrt{\sigma_j\sigma_i}}{\sigma_j - \sigma_i} + O(q) &\sim &&\sum_{i \in \{-,b\}} t_j t_i \lambda_i \frac{S_{cyl}(z_j, z_i)}{\sqrt{\ud z_j \ud z_i}} + O(q), \nonumber\\
  \tilde{\lambda}_j =& \, t_j \tilde{\lambda}_0 + \sum_{i \in \{+\}} t_j t_i \tilde{\lambda}_i \frac{S_1(z_j, z_i; \tau)}{\sqrt{\ud z_j} \sqrt{\ud z_i}} &\rightarrow && t_j \tilde{\lambda}_0 + \sum_{i \in \{+\}} \frac{t_j t_i \tilde{\lambda}_i \sqrt{\sigma_j\sigma_i}}{\sigma_j - \sigma_i} + O(q) =  \label{ScatteringEquations}\\
  &\frac{t_j t_a \tilde{\lambda}_a}{- \sqrt{\sigma_a}}  + \sum_{i \in \{+\}} \frac{t_j t_i \tilde{\lambda}_i \sqrt{\sigma_j\sigma_i}}{\sigma_j - \sigma_i} + O(q) &\sim &&\sum_{i \in \{+,a\}} t_j t_i \tilde{\lambda}_i  \frac{S_{cyl}(z_j, z_i)}{\sqrt{\ud z_j \ud z_i}} + O(q), \nonumber\\ \nonumber
  \end{align}
  where we identify $\lambda_0$ with $\lambda_b = -\lambda_a$ up to a scale $-\frac{t_b}{\sqrt{\sigma_b}}$ and $\tilde{\lambda}_0$ with $\tilde{\lambda}_a = \tilde{\lambda}_b$ up to a scale $-\frac{t_a}{\sqrt{\sigma_a}}$, together with the understanding that for $q \!\to\! 0 \,\, \sigma_a$ is infinitely far away from the punctures with negative helicity and $\sigma_b$ infinitely far away from the punctures with positive helicity. This way we end up with modified scattering equations for the external particles that involve the new particles on the right hand side. We add scattering equations for the new particles themselves by making the replacement
  \begin{equation*}
  \begin{aligned}
  &\ud^2 \!\lambda_0 \, \ud^2 \!\tilde{\lambda}_0 \, \delta^4\!(\sum_{i = 1}^n \tilde{\lambda}_i \lambda_i) &\rightarrow &&\ud^2 \!\lambda_a \, \ud^2 \!\tilde{\lambda}_b \, \delta^2\!(\lambda_a \!-\! t_a \lambda(z_a))  \,\delta^2\!(\tilde{\lambda}_b \!-\! t_b \tilde{\lambda}(z_b)) \,.
  \end{aligned}
  \end{equation*}
  This can be done because now we have scattering equations for all particles (including the additional $a$ and $b$ particles) and on their support a delta function expressing momentum conservation can be factored out. This momentum conservation is for all particles, but reduces to only external particles as well because of the condition $\tilde{\lambda}_a \lambda_a + \tilde{\lambda}_b \lambda_b = 0$. Based on \eqref{Rfactorization} and above identifications for $\lambda_0$ and $\tilde{\lambda}_0$ we replace in $\Phi_1$ every term $-t_r [r0]$ with $-t_r t_a [\tilde{\lambda}_r \tilde{\lambda}_b] \sqrt{q}\frac{S_{cyl}(z_r - z_a)}{\sqrt{\ud z_r \ud z_a}}$ and in $\tilde{\Phi}_1$ every term $-t_l \braket{l0}$ with $-t_l t_b \braket{\lambda_l \lambda_a} \sqrt{q} \frac{S_{cyl}(z_l - z_b)}{\sqrt{\ud z_l \ud z_b}}$ in cylinder coordinates. Like in the NS sector case we get, this time on the cylinder and including the factor $t_a[\tilde{\lambda}_b \tilde{\lambda}(z_a)] t_b\!\braket{\lambda_a \lambda(z_b)}$ from \eqref{OnShellVertices} under the support of the scattering equations:
 \begin{equation}
 \begin{split}
 &\mathcal{M} \!=\!\! \int \!\! \frac{\ud^2 \!\lambda_a  \ud^2 \!\tilde{\lambda}_b}{\mathrm{vol} \mathbb{C}^{*}}\int \!\! \frac{1}{\mathrm{vol GL}(2, \mathbb{C})} \!\! \prod_{j  \in \{+,a\} \atop{\cup \{-,b\}}} \!\!\frac{\ud t_j \ud z_j} {t_j^3} \!\!
 \prod_{k \in \{+,a\}} \!\!\! \delta^2\!(\lambda_k \! - \! t_k \lambda(z_k)) \!\!\! \prod_{l \in \{-,b\}} \!\!\! \delta^2\!(\tilde{\lambda}_l \! - \! t_l \tilde{\lambda}(z_l))\\
 &\!\!\left[ \!\Bigl(\!\!\sum_{r \in \{+\}} \!\! t_r t_a [\tilde{\lambda}_r \tilde{\lambda}_b] \frac{S_{cyl}(z_r \!\!-\!\! z_a)}{\sqrt{\ud z_r \ud z_a}}\Bigr)^{\!2} \!\!\!\!\prod_{s \in \{+\}} \!\!\!\!\Phi_{ss}^{tr} \mathrm{det}^{\prime} \Phi^{tr} \right]
 \!\!\left[  \!\Bigl(\!\!\sum_{l \in \{-\}} \!\! t_l t_b \braket{\lambda_l \lambda_a} \!\frac{S_{cyl}(z_l \!\!-\!\! z_b)}{\sqrt{\ud z_l \ud z_b}} \Bigr)^{\!2} \!\!\!\!\prod_{k \in \{-\}} \!\!\!\!\tilde{\Phi}_{kk}^{tr}  \mathrm{det}^{\prime} \tilde{\Phi}^{tr} \right]\!\!,
 \end{split}
 \label{RNonSepFactorization}
 \end{equation}
  where $\Phi^{tr} = \Phi^{tree}$ and $\tilde{\Phi}^{tr} = \tilde{\Phi}^{tree}$ arise from $\Phi_1$ and $\tilde{\Phi}_1$ by replacing all propagators with a propagator for genus 0, respectively, and by deleting the terms containing the zero modes $\lambda_0$ and $\tilde{\lambda}_0$.
 Particle exchange with opposite momenta is then guaranteed by inserting
 \begin{equation*}
 1 = \int \ud^2 \!\tilde{\lambda}_a \ud^2 \!\lambda_b \,\, \delta^2(\lambda_b \!+\! \lambda_a)\, \delta^2(\tilde{\lambda}_a \!-\! \tilde{\lambda}_b) \,.
 \end{equation*}
 
 This should coincide with a single unitary cut of a tree amplitude with $n+2$ particles like in \eqref{single_cut}, on the cylinder
 \begin{align}
 &\mathcal{M}_{k,n}^{1-loop} = \! \int \!\! \frac{\ud^2 \lambda_0 \ud^2 \tilde{\lambda}_0}{\mathrm{vol} \mathbb{C}^{*}}\int \frac{\ud q}{q^2} \mathcal{M}_{k+1,n+2}^{tree}(q; \{\lambda_i, \tilde{\lambda}_i ; \lambda_0, \tilde{\lambda}_0 \}_{i \in \{-\}} ; \! \{\lambda_j, \tilde{\lambda}_j ; -\!\lambda_0, \tilde{\lambda}_0 \}_{j \in \{+\}}) \,,\nonumber\\
 &\mathcal{M}_{k,n}^{tree} \!=\! \int \!\! \frac{ \prod_{\{+\}} \!\!\Phi_{ss}^{tr} \mathrm{det}^{\prime} \Phi^{tr} \prod_{\{-\}} \!\!\tilde{\Phi}_{ll}^{tr}  \mathrm{det}^{\prime} \tilde{\Phi}^{tr}}{\mathrm{vol GL}(2, \mathbb{C})} \!\!\!\! \prod_{j  \in \{+\} \atop{\cup \{-\}}} \!\!\!\!\frac{\ud t_j \ud z_j} {t_j^3} \!
 \!\!\!\prod_{s \in \{+\}} \!\!\! \delta^2\!(\lambda_s \! - \! t_s \lambda(z_s)) \!\!\!\! \prod_{l \in \{-\}} \!\!\! \delta^2\!(\tilde{\lambda}_l \! - \! t_l \tilde{\lambda}(z_l)), \label{RTree}
 \end{align}
 where the subscript $k,\!n$ means $k$ particles with negative helicity and $n-k$ particles with positive helicity, and the $q$-dependence of $\mathcal{M}^{tree}$ refers to the modeling equation \eqref{Rfactorization}, again with the interpretation that in the limit $q \rightarrow 0$ the puncture locations become infinitely separated from the puncture of the loop particle belonging to the other helicity set, like in \eqref{ScatteringEquations}. Repeating the procedure done above for the even spin structure case it is straightforward to see that in the scaling limit $\mathcal{M}_{k,n}^{1-loop}$ indeed agrees with \eqref{RNonSepFactorization}:
 the amplitude becomes
 \begin{equation}
 \begin{split}
 \mathcal{M}_{k,n}^{1-loop} \!=\!\! \int \!\! \frac{\ud^2 \!\lambda_0  \ud^2 \!\tilde{\lambda}_0}{\mathrm{vol} \mathbb{C}^{*}}&\int \!\! \frac{1}{\mathrm{vol GL}(2, \mathbb{C})} \!\! \prod_{j  \in \{+,a\} \atop{\cup \{-,b\}}} \!\! \frac{\ud t_j \ud z_j}{t_j^3} \!\!
 \prod_{s \in \{+,a\}} \!\!\! \delta^2\!(\lambda_s \! - \! t_s\lambda(z_s)) \!\!\! \prod_{l \in \{-,b\}} \!\!\! \delta^2\!(\tilde{\lambda}_l \! - \! t_l\tilde{\lambda}(z_l))\\
 &\lim_{q \to 0} \frac{1}{q} \left( \prod_{s \in \{+,a\}} \!\!\Phi_{n-k+1\, ss}^{tr}\mathrm{det}^{\prime} \Phi_{n-k+1}^{tr} \,  \prod_{l \in \{-,b\}} \!\!\tilde{\Phi}_{k+1\,ll}^{tr} \mathrm{det}^{\prime} \tilde{\Phi}_{k+1}^{tr}\right)
 \end{split}
 \label{Rsingle_cut}
 \end{equation}
 with
 \begin{align}
 &\lim_{q \to 0} \frac{\prod_{\{+,a\}} \!\!\Phi_{n-k+1\, ss}^{tr}}{\sqrt{q}} \mathrm{det}^{\prime} \Phi_{n-k+1}^{tr} \!\! =  \!\!\! \lim_{z_b - z_j \to z_b \atop{(\forall j \in \{+\})}} \!\Bigl(\!\sum_{i \in \{+\} } t_a t_i [0 \, i] \frac{S_{cyl}(z_a \!\!-\!\! z_i)}{\sqrt{\ud z_a \ud z_i}} \Bigr)^{\!2} \!\!\!\prod_{s \in \{+\}} \!\!\!\!\Phi_{n-k\,ss}^{tr} \, \mathrm{det}^{\prime} \Phi_{n-k}^{tr}  \,, \nonumber\\
 &\lim_{q \to 0} \frac{\prod_{\{-,b\}} \!\!\tilde{\Phi}_{k+1\,ll}}{\sqrt{q}} \mathrm{det}^{\prime} \tilde{\Phi}_{k+1}^{tr} = \!\!\! \lim_{z_a - z_m \to  z_a \atop{(\forall m \in \{-\})}}  \!\Bigl(\!\sum_{l \in \{-\}} t_b t_l \braket{0 \, l} \frac{S_{cyl}(z_b \!\!-\!\! z_l)}{\sqrt{\ud z_b \ud z_l}}  \Bigr)^{\!2} \!\!\!\prod_{l \in \{-\}} \!\!\!\!\tilde{\Phi}_{k\,ll}^{tr} \, \mathrm{det}^{\prime} {\tilde\Phi}_k^{tr} ,
 \label{Rsingle_cut_matrices}
 \end{align}
 where $\Phi_{n-k}^{tr}$ and $\tilde{\Phi}_{k}^{tr}$ are the n-particle tree amplitude matrices obtained from $\Phi_{n-k+1}^{tr}$ and $\tilde{\Phi}_{k+1}^{tr}$, as in the even spin structure case, by deleting all entries referring to $\sigma_a$ and $\sigma_b$ and deleting the row and column containing only zero entries, respectively. This indeed matches \eqref{RNonSepFactorization} when taking the limit. Again, other loop particles can be added by inserting the appropriate fixed vertex operators, in \eqref{OnShellVertices} and in the $\prod_{\{\pm\}}$ factors of \eqref{RTree}.\\
 
 Therefore, the single unitary cut of an $n+2$ particle tree amplitude, when performed in the scaling limit where in the scattering equations the remote location of the loop particle becomes infinitely separated from all punctures belonging to the other helicity set, can indeed be identified with the one-loop amplitude in the non-separating degeneration limit. The recipe to calculate an $n$-particle one-loop amplitude in the non-separating degeneration limit from an $n+2$-particle tree amplitude (one additional particle per helicity) is to solve the scattering equations for $n+2$ particles and plugging the result into equations \eqref{single_cut_matrices} and \eqref{single_cut} in the NS sector and into \eqref{Rsingle_cut_matrices} and \eqref{Rsingle_cut} in the R sector. In appendix \ref{4GravOneLoop} it is shown that this recipe leads to the expected one-loop amplitude for four external gravitons in the NS sector.\\

\section{String Field Theory}
  \label{SFT}
  In this section we want to sketch the construction of a String Field Theory (SFT) for our model, following the exposition in \cite{Reid:2017} and \cite{Lacroix:2017}. It will look similar to the SFT of the ambitwistor string in 10 dimensions \cite{Reid:2017}, but it is more complicated because of the proliferation of ghost fields.\\
      
  We have to decide in which picture numbers and ghost numbers to work. In our model the bosonic antighost-ghost pairs do not switch boundary conditions and have conformal weight 1 and 0. When doing the bosonization of one of these ghosts (see appendix \ref{StringProperties}) the operator $e^{- \frac{1}{2} \varphi}$ that switches picture number 0 to $-\frac{1}{2}$ has conformal weight $\frac{1}{8}$ which adds up to 4 with all 32 bosonic ghosts and would give trouble for coming up with a proper conformal dimension of vertex operators. On the other hand, the operator $e^{- \varphi}$ that switches picture number 0 to -1 has conformal weight 0. Therefore, we will assume that vertex operators in the R sector always create cuts in pairs. This way, vertex operators in both (NS and R) sectors are in picture number -1 for every bosonic ghost. Regarding the ghost number we take all vertex operators for on-shell external states in ghost number 1 for every fermionic ghost. For later use,\\
  by $\mathcal{H}$ we denote the small Hilbert space of all CFT states (see appendix \ref{StringProperties}), \\
  by $\mathcal{H}_0$ the subspace of states in picture number -1 for every bosonic ghost, \\
  by $\widetilde{\mathcal{H}}_0$ the subspace of states in picture number 0 for every bosonic ghost, \\
  by $\mathcal{H}_1$ the subspace of $\mathcal{H}_0$ with ghost number 1 for every fermionic ghost, and\\
  by $\widetilde{\mathcal{H}}_1$ the subspace of $\widetilde{\mathcal{H}}_0$ with ghost number 0 for every fermionic ghost except for the $c$ ghost with ghost number 1. 
  
  Like in \cite{Lacroix:2017} we denote by $\mathcal{M}_{g,m,n}$  the $(3g - 3 + m + n)$ dimensional moduli space of genus $g$ Riemann surfaces $\Sigma_{g,m,n}$ with $m$ NS and $n$ R punctures. In order to define off-shell amplitudes we also define the space $\tilde{\mathcal{P}}_{g,m,n}$ with the structure of a fiber bundle, whose base is $\mathcal{M}_{g,m,n}$ and whose fiber is parametrized by the possible choices of a local coordinate system around each puncture and the possible choices of PCO locations on the Riemann surface. For $2g - 2 + m + n > 0$, the Riemann surface $\Sigma_{g,m,n}$ can be regarded as a union of $m + n$ disks $\{D_{\alpha}\}$, one around each puncture, and $2g - 2 + m + n$ spheres $\{S_i\}$, each with 3 holes, joined along $3g - 3 + 2(m + n)$ circles $\{C_s\}$\footnote{For $2g - 2 + m + n \le 0$ a single sphere with up to two holes is sufficient.}. By $\{u_i, y_{\alpha}\}$ we denote a coordinate system for $\tilde{\mathcal{P}}_{g,m,n}$ where the $u_i$ parametrize the Riemann surface and local coordinates around the punctures and the $y_{\alpha}$ are the locations of the $g - 1 + m + n$ PCOs, in analogy to \cite{Lacroix:2017}. Associated with infinitesimal motions in $\tilde{\mathcal{P}}_{g,m,n}$ are tangent vectors $\frac{\partial}{\partial y_{\alpha}}$ for changes in the PCO locations and  $\frac{\partial}{\partial u_i}$ for deformations of coordinate transition functions $F_s$ on overlap circles $\{C_s\}$. We define, in analogy to \cite{Reid:2017} and \cite{Lacroix:2017},
  \begin{equation*}
  \begin{split}
  & \boldsymbol{B}[\frac{\partial}{\partial u_i}] = \sum_s \oint_{C_s} \frac{\partial F_s}{\partial u_i} \ud \sigma_s b(\sigma_s),\\
  & \boldsymbol{H}[\frac{\partial}{\partial u_i}] = \sum_s \oint_{C_s} \frac{\partial F_s}{\partial u_i} \ud \sigma_s \vec{h}(\sigma_s) \cdot
                                                                           \bar{\delta}\left( \sum_s \oint_{C_s} \frac{\partial F_s}{\partial u_i} \ud \sigma_s \{Q, \vec{h}(\sigma_s)\} \right),\\
  & \boldsymbol{F}[\frac{\partial}{\partial u_i}] =  \prod_{i,j, \dot{\alpha},\alpha} \left( \sum_s \oint_{C_s} \frac{\partial F_s}{\partial u_i} \ud \sigma_s f_{i j \dot{\alpha} \alpha}(\sigma_s) \right) \bar{\delta}\left( \sum_s \oint_{C_s} \frac{\partial F_s}{\partial u_i} \ud \sigma_s \{Q, f_{i j \dot{\alpha} \alpha}(\sigma_s)\} \right),\\
  & \tilde{\boldsymbol{B}}[\frac{\partial}{\partial u_i}]  = \boldsymbol{H}[\frac{\partial}{\partial u_i}] \boldsymbol{F}[\frac{\partial}{\partial u_i}],
  \end{split}
  \end{equation*}
  and use this definition to construct on $\tilde{\mathcal{P}}_{g,m,n}$ a $p$-form $\Omega_p^{(g,m,n)}(\{K_i\}, \{L_j\})$ for (potentially off-shell) $m$ NS and $n$ R vertex operators $\{K_i\}$ and $\{L_j\}$ inserted at the punctures corresponding to states in $\mathcal{H}_0$, to be integrated over a $p$-dimensional subspace of $\tilde{\mathcal{P}}_{g,m,n}$. It is defined by contracting it with $p$ arbitrary tangent vectors of $\tilde{\mathcal{P}}_{g,m,n}$:
  \begin{equation}
  \begin{split}
  &\Omega_p^{(g,m,n)}(\{K_i\}, \{L_j\})[\frac{\partial}{\partial u_{i_1}},\cdots,\frac{\partial}{\partial u_{i_k}},\frac{\partial}{\partial y_{\alpha_{k+1}}},\cdots,\frac{\partial}{\partial y_{\alpha_p}}] = (-2 \pi i)^{-(3g - 3 + m + n)} \\
  &\left< \prod_{j=1}^k \!\! \boldsymbol{B}[\frac{\partial}{\partial u_{i_j}}]  \!\!\prod_{i=1 \atop{i \in \{i_1, \cdots , i_k\}}}^{g - 1 + m + n} \!\!\!\!\! \tilde{\boldsymbol{B}}[\frac{\partial}{\partial u_i}] \!\prod_{q=k+1}^p \!\!\left[\sum^{32}_{l \in \{{\dot{\beta} \beta \atop {rst}}\}} \!\!\!\!\frac{- \partial\xi_{\ell}(y_{\alpha_q}\!)}{\chi_{\ell}(y_{\alpha_q}\!)}\!\!\right]\!
  \!\!\!\!\!\!\!\!\!\prod_{\alpha = 1}^{\,\,\,\,\,\,\,\,\,\,\,g - 1 + m + n}\!\!\!\!\!\!\!\!\!\! \chi(y_{\alpha}) \! K_1 \cdots K_m, L_1, \cdots, L_n \!\right>_{\!\!\!\!\!\!\Sigma_{g,m,n}} \!\!\!\!\!\!,
  \end{split}
  \label{OffShell}
  \end{equation}
  where $\chi(y_{\alpha})$ and $\chi_{\ell}(y_{\alpha})$ are PCOs \eqref{PictureNumber} and dividing by $\chi_{\ell}(y_{\alpha})$ means skipping this factor in $\chi(y_{\alpha})$ at the same location. This expression is evaluated at some specific point of $\tilde{\mathcal{P}}_{g,m,n}$, depending on the choice of local coordinates and PCO locations. $\braket{\cdots}_{\Sigma_{g,m,n}}$ stands for the correlation function on the Riemann surface $\Sigma_{g,m,n}$. Because of ghost number conservation, $\Omega_p^{(g,m,n)}(\{K_i\}, \{L_j\})$ does not vanish only when the total ghost number inside the correlation is equal to $22(1 - g)$. On the other hand, picture numbers are conserved by design. 
  
  We now define off-shell amplitudes as
  \begin{equation*}
  \int_{S_{g,m,n}} \Omega_{3g - 3 + m + n}^{(g,m,n)}(K_1, \cdots, K_m, L_1, \cdots, L_n) \,,
  \end{equation*}
  where $S_{g,m,n}$ is a section of $\tilde{\mathcal{P}}_{g,m,n}$ \footnote{When $g = 0$, $\Omega_{3g - 3 + m + n}^{(g,m,n)}$ should be replaced with $\Omega_{- 1 + m + n}^{(0,m,n)}$ and the product $\prod_{j=1}^k \! \boldsymbol{B}$ in the correlation of \eqref{OffShell} with $\prod_{j=3}^k \! \boldsymbol{B}$. Henceforth, we assume this replacement is done implicitly in all formulas containing a sum over $\Omega_{3g - 3 + m + n}^{(g,m,n)}$ starting at $g = 0$.}. Without proof we assume that these amplitudes, when the external states are on-shell, reduce to the usual on-shell amplitudes and are independent of the choice of the section $S_{g,m,n}$. We also assume that these amplitudes share the analogous properties of the off-shell amplitudes of the superstring defined in \cite{Lacroix:2017}, required to define a consistent string field theory. The main difference between the off-shell amplitudes of our model and the ones of the superstring are the non-local operators $\tilde{\boldsymbol{B}}$ (a complication) and the fact that all vertex operators are in picture number -1 for both the NS and R sector (a simplification). In \cite{Reid:2017} similar non-local operators are introduced for the ambitwistor string field theory, but the situation there looks simpler because the only fermionic ghost system beyond the $b\!-\!c$ system has the same conformal weights.\\
  
  As in \cite{Lacroix:2017} we denote by $\bar{\mathcal{R}}_{g,m,n}$ the section segments of the elementary vertices at genus g with $m$ and $n$ external legs in the NS and R sector, respectively, and define
  \begin{equation}
  \{\!\!\{K_1, \cdots, K_m, L_1, \cdots, L_n\}\!\!\} = \sum_{g = 0}^{\infty} g_s^g \int_{\bar{\mathcal{R}}_{g,m,n}} \Omega_{3g - 3 + m + n}^{(g,m,n)}(K_1, \cdots, K_m, L_1, \cdots, L_n) \,,
  \label{MultiLinear}
  \end{equation} 
  where $g_s$ is the string coupling. The string field action then becomes
  \begin{equation*}
  S = \frac{1}{g_s} \left[ \frac{1}{2} \bra{\tilde{\Psi}} Q \ket{\Psi} + \sum_{n=1}^{\infty} \{\!\!\{ \Psi^n \}\!\!\} \right],
  \end{equation*}
  where $\ket{\Psi}$ is an element of $\mathcal{H}_0$, $\ket{\tilde{\Psi}}$ one of $\widetilde{\mathcal{H}}_0$, and $\bra{\tilde{\Psi}}$ is the BPZ conjugate of  $\ket{\tilde{\Psi}}$.
  Instead of $\bar{\mathcal{R}}_{g,m,n}$ one can consider the section segments $\mathcal{R}_{g,m,n}$ of the 1PI elementary vertices at genus g with $m$ NS and $n$ R external legs, constructed using plumbing of only separated Riemann surfaces, define the multilinear functions $\{K_1, \cdots, K_m, L_1, \cdots, L_n\}$ as in \eqref{MultiLinear}, but with $\mathcal{R}_{g,m,n}$ replacing $\bar{\mathcal{R}}_{g,m,n}$, and then define the 1PI action
  \begin{equation*}
  S = \frac{1}{g_s} \left[ \frac{1}{2} \bra{\tilde{\Psi}} Q \ket{\Psi} + \sum_{n=1}^{\infty} \{ \Psi^n \} \right],
  \end{equation*}
  which leads to the classical action as restriction of $\{ \Psi^n \}$ to genus 0:
  \begin{equation*}
  S_{cl} = \frac{1}{g_s}  \left[ \frac{1}{2} \bra{\tilde{\Psi}} Q \ket{\Psi} + \sum_{n=3}^{\infty} \{ \Psi^n \}_0 \right],
  \end{equation*}
  where now $\ket{\Psi}$ can be taken as an element of $\mathcal{H}_1$ and $\ket{\tilde{\Psi}}$ one of $\widetilde{\mathcal{H}}_1$.\\
  
  Again, it is assumed that the multilinear functions $\{\!\!\{ \cdots \}\!\!\}$ and $\{ \cdots \}$ and the actions have the right properties to define a consistent string field theory including satisfying the quantum BV master equation. To prove or disprove this, is beyond the scope of this paper. It would be interesting to compare this action with similarly looking twistor actions introduced in the literature (for a review see \cite{Adamo_2:2013}).\\

\section{Summary and Outlook}
In this paper we have investigated some properties of the anomaly free twistor string in more detail. The spectrum in the NS sector has a graviton leading to proper gravitational amplitudes. In the R sector the spectrum is similar to the one for conformal supergravity of the Berkovits-Witten twistor string, enhanced by being multiplied with an SU(2) representation of the bi-twistor, but without the states thought of spoiling unitarity and without the requirement to be in an adjoint representation of a super multiplet. Part of the spectrum for spin $\frac{1}{2}$ fermions can be chosen to look like 3 generations of fermionic content of the Pati-Salam model \cite{Pati:1974}. All helicity states have double occurrence, reflecting the symmetry between matter fields and their duals. Scattering amplitudes with external gravitons and vector bosons have been calculated in the NS sector and they match the EYM amplitudes of \cite{Adamo:2015}. All tree and one-loop amplitudes of \cite{Kunz:2020} have been confirmed to exhibit proper unitary factorization. In appendix \ref{4GravOneLoop} the one-loop amplitude with four external gravitons has been explicitly calculated in the non-separating degeneration limit and verified to replicate the expected result. Finally, the construction of a string field theory similar to the one for the conventional ambitwistor string in 10 dimensions \cite{Reid:2017, Lacroix:2017} has been proposed.

Among the many open issues we only list a few  obvious ones and one farfetched one:
\begin{itemize}
\item The SFT construction proposed in section \ref{SFT} needs to be put on rigorous mathematical ground with many more details worked out. In particular, consistency, UV finiteness, and unitarity need to be established. That the model has proper factorization properties and modular invariance up to the one-loop level is a promising starting point. 
\item If the SFT exists as a consistent theory, does the model have physical significance? The classical limit would need to be understood more deeply. In particular, can a connection with general relativity be established? The fact that it leads to expected EYM scattering amplitudes is a hopeful sign. 
\item Usually, black hole physics is used as a playground for quantum gravity theories. What does the model have to say about black hole properties like the information paradox phenomenon? Because the target space is twistor space, black holes might need to be modeled in that space. One typical way is to use non-Hausdorff reduced twistor spaces \cite{Fletcher:1988, Fletcher:1989} although it does not seem easily feasible to consider the twistor string on such unusual background because of the implicit connection to spacetime.
\item A final, rather speculative item: Is there a connection between the SFT (if it exists) and loop quantum gravity (LQG)? The twistorial interpretation of LQG \cite{Speziale:2012} works with a pair of twistors, together with a simplicity constraint and an area matching condition. The former looks like a gauged complex incidence relation and the latter like a scaling symmetry between the two twistors, which both could be mapped into our model. But is there more to it than just this correspondence?
\end{itemize}
\ \\

\appendix
\section{Basic String Properties} 
\label{StringProperties}
  The twistor fields $(\lambda_{i\alpha},\! \mu_i^{\dot{\alpha}} ), (\tilde{\mu}_i^{\alpha},\!  \tilde{\lambda}_{i\dot{\alpha}}) (i \!=\! 1,\!2)$ and fermionic fields $(\phi_{i\alpha},\!\psi_i^{\dot{\alpha}}),\! (\tilde{\psi}_i^{\alpha}, \tilde{\phi}_{i\dot{\alpha}})$ $(i \!=\! 1,\!2)$ have the mode expansion:
  \begin{align*}
  \lambda_{i\alpha} & = \sum_n \! \lambda_{i\alpha\,n} z^{-n-\frac{1}{2}}, & \mu_i^{\dot{\alpha}} & = \sum_n \! \mu_{i\,n}^{\dot{\alpha}} z^{-n-\frac{1}{2}},  
  & \tilde{\mu}_i^{\alpha} & = \sum_n \! \tilde{\mu}_{i\,n}^{\alpha} z^{-n-\frac{1}{2}}, & \tilde{\lambda}_{i\dot{\alpha}} & = \sum_n \! \tilde{\lambda}_{i\dot{\alpha}\,n} z^{-n-\frac{1}{2}}, &\\
  \phi_{i\alpha} & = \sum_n \! \phi_{i\alpha\,n} z^{-n-\frac{1}{2}}, & \psi_i^{\dot{\alpha}} & = \sum_n \! \psi_{i\,n}^{\dot{\alpha}} z^{-n-\frac{1}{2}},
  & \tilde{\psi}_i^{\alpha}  & = \sum_n \! \tilde{\psi}_{i\,n}^{\alpha}  z^{-n-\frac{1}{2}}, & \tilde{\phi}_{i\dot{\alpha}} & = \sum_n \! \tilde{\phi}_{i\dot{\alpha}\,n} z^{-n-\frac{1}{2}},  &
  \end{align*}
  with $n \, \epsilon \, \mathbb{Z}$ in the R sector and $n \, \epsilon \, \mathbb{Z} + \frac{1}{2}$ in the NS sector. The ghost system consists of 20 fermionic $(b,c), (\vec{h}, \vec{g}), (f_{ij\, \dot{\alpha} \alpha}, e^{\dot{\alpha} \alpha}_{ij})$ and 32 bosonic $(\beta_{kij \, \alpha \dot{\alpha}}, \gamma^{\alpha \dot{\alpha}}_{kij})$ anti-ghost and ghost fields. $(b,c)$ has conformal weight $(2,-1)$ and all other anti-ghost ghost pairs have conformal weight $(1,0)$. Henceforth we will sometimes drop the $\alpha$ and $\dot{\alpha}$ indices when their presence is implicitly clear.
  The $(\beta_{kij \, }, \gamma_{kij})$ can be bosonized as
  \begin{align*}
  \gamma_{k i j} & = :\! \eta_{k i j} \, e^{\varphi_{k i j}} \!:, & \beta_{k i j} & = :\! \partial \xi_{k i j} \, e^{-\varphi_{k i j}} \!:, &\delta(\gamma_{k i j}) & = :\! e^{-\varphi_{k i j}} \!:, & \delta(\beta_{k i j}) & = :\! e^{\varphi_{k i j}} \!:, &
  \end{align*}
  where $\eta_{k i j}, \xi_{k i j}$ are fermions of conformal weights 1 and 0, respectively, and $\varphi_{k i j}$ is a scalar with background charge $- \! \frac{1}{2}$.
  
  The ghost fields have the mode expansion
  \begin{align*}
  c & = \sum_n \! c_n z^{-n+1}, & b & = \sum_n b_n z^{-n-2}, &
  \vec{g} & = \sum_n \! \vec{g}_n z^{-n}, & \vec{h} & = \sum_n \vec{h}_n z^{-n-1}, &\\
  e_{i j} & = \sum_n \! e_{i j n} z^{-n}, & f_{i j} & = \sum_n f_{i j n} z^{-n-1}, &
  \gamma_{k i j} & = \sum_n \! \gamma_{k i j n} z^{-n}, & \beta_{k i j} & = \sum_n \beta_{k i j n} z^{-n-1},&\\
  \eta_{k i j} & = \sum_n \! \eta_{k i j n} z^{-n-1}, & \xi_{k i j} & = \sum_n \xi_{k i j n} z^{-n},&
  \end{align*}
  with $n \, \epsilon \, \mathbb{Z}$.
 
 The vacuum is annihilated for $n > 0$ by
 \begin{align*}
 \lambda_{i\,n} \ket{0} &= 0, &\mu_{i \, n} \ket{0} &= 0, &\tilde{\lambda}_{i\,n\!-\!\frac{1}{2}}  \ket{0} &= 0, &\tilde{\mu}_{i \, n\!-\!\frac{1}{2}} \ket{0} &= 0, &\\
 \phi_{i\,n} \ket{0} &= 0, &\psi_{i \, n} \ket{0}  &= 0, &\tilde{\phi}_{i\,n\!-\!\frac{1}{2}}  \ket{0} &= 0, &\tilde{\psi}_{i \, n\!-\!\frac{1}{2}} \ket{0} &= 0, &\\
  c_{n\!+\!1} \ket{0} &= 0, &b_{n\!-\!2} \ket{0} &= 0, &\vec{g}_n \ket{0} &= 0, &\vec{h}_{n\!-\!1}  \ket{0} &= 0, &\\
  e_{i j \, n} \ket{0} &= 0, &f_{i j \, n\!-\!1} \ket{0} &= 0, &\gamma_{k i j \, n} \ket{0} &= 0, &\beta_{k i j \, n\!-\!1} \ket{0} &= 0. &
  \end{align*}
  We shall be working in the so called 'small Hilbert space' (denoted by $\mathcal{H}$) with the zero modes $\xi_{k i j \, 0}$ of the $\xi_{k i j}$ fields missing from the spectrum. This means that the states are annihilated by $\eta_{k i j \, 0}$s.
  The nilpotent BRST operator is $Q= \oint \ud z j_B(z)$, where the BRST current $j_B(z)$ is (with implicit summation over repeated indices)
  \begin{equation*}
  \begin{split}
  j_B(z) &= c(z) \left(T(z) + \frac{1}{2} T_{b,c}(z) + T_{f,e}(z) + T_{\beta, \gamma}(z) + T_{h,g}(z) - \frac{1}{z^2}\right)\\
  & + e_{ij}^{\dot{\alpha}_i \alpha_j}(z) F_{ij\, \dot{\alpha}_i \alpha_j}(z) + \gamma_{1ij}^{\alpha_i \dot{\beta}_j}(z) G_{1ij\, \alpha_i \dot{\beta}_j}(z) + \gamma_{2ij}^{\dot{\alpha}_i \beta_j}(z) G_{2ij\, \dot{\alpha}_i \beta_j}(z) + \vec{g}(z) \cdot \vec{H}(z)\\
  & + \frac{i}{2} \, :\! \vec{g}(z) \cdot (\vec{g}(z) \times \vec{h}(z)) \!: + \vec{g}(z) \cdot \left(\vec{H}_{f,e}(z) + \vec{H}_{\beta, \gamma}(z)\right) \,,
  \end{split}
  \end{equation*}
  where $\oint$ is normalized to $\oint \frac{\ud z}{z} = 1$ and the various currents are defined as\footnote{On a side note, the gauged action can be regained from the energy-momentum current as $S \!=\! -\frac{1}{2\pi} \!\int\! \ud^2z [ T(z) + T_{b,c}(z) + T_{f,e}(z) + T_{\beta, \gamma}(z) + T_{h,g}(z) ]_{\partial \rightarrow \bar{\partial}}$.}
  \begin{align}
  &T(z) = \frac{1}{2} : \! \left[ \braket{\lambda_i(z) (\partial \tilde{\mu}_i)(z)} - \braket{\tilde{\mu}_i(z) (\partial \lambda_i)(z)} + [\mu_i(z) (\partial \tilde{\lambda}_i)(z)] - [\tilde{\lambda}_i(z) (\partial \mu_i)(z)]  \right] \!: \nonumber\\
  &\,\,\,\,\,\,\,\,\,\,\,\,- \frac{1}{2} : \! \left[ \braket{\phi_i(z) (\partial \tilde{\psi}_i)(z)} + \braket{\tilde{\psi}_i(z) (\partial \phi_i)(z)} +  [\psi_i(z) (\partial \tilde{\phi}_i)(z)] +  [\tilde{\phi}_i(z) (\partial \psi_i)(z)]  \right] \!: \,, \nonumber\\
  &T_{b,c}(z) = - :\! \left( b(z) \partial c(z) + \partial(b \, c)(z) \right) \!: \,, \nonumber\\
  &T_{f,e}(z) = - :\! f_{ij \, \dot{\alpha}_i \alpha_j}(z) (\partial e_{ij}^{\dot{\alpha}_i \alpha_j})(z) \!: \,, \nonumber\\
  &T_{\beta,\gamma}(z) = - :\! \left(\beta_{1ij \, \alpha_i \dot{\beta}_j}(z) (\partial \gamma_{1ij}^{\alpha_i \dot{\beta}_j})(z) + \beta_{2ij \, \dot{\alpha}_i \beta_j}(z) (\partial \gamma_{2ij}^{\dot{\alpha}_i \beta_j})(z) \right) \!: \nonumber\\
  &\,\,\,\,\,\,\,\,\,\,\,\,\,\,\,\,\, = - :\!\eta_{kij}(z) (\partial \xi_{kij})(z) \!: - \frac{1}{2} :\! \left((\partial \varphi_{kij})(z)^2 + (\partial^2 \varphi_{kij})(z)\right) \!: \,,\nonumber\\
  &T_{h,g}(z) = - :\! \vec{h}(z) \cdot (\partial \vec{g})(z) \!: \,, \nonumber\\
  &F_{ij\, \dot{\alpha}_i \alpha_j}(z) = - \tilde{\lambda}_{i\dot{\alpha}_i}(z) \lambda_{j \alpha_j}(z) \,, \nonumber\\  
  &G_{1i j\,\alpha_i \dot{\beta}_j}(z) = \lambda_{i \alpha_i}(z) \tilde\phi_{j \dot{\beta}_j}(z) \,, \nonumber\\ 
  &G_{2i j\,\dot{\alpha}_i \beta_j}(z) = - \tilde\lambda_{i \dot{\alpha}_i}(z) \phi_{j \beta_j}(z) \,, \nonumber\\
  &\vec{H}(z) = - :\! \left( \braket{(\tilde{\mu}_1(z) \tilde{\mu}_2(z)) \vec{\tau} \binom{\lambda_1(z)}{\lambda_2(z)}} + [(\tilde{\lambda}_1(z) \tilde{\lambda}_2(z)) \vec{\tau} \binom{\mu_1(z)}{\mu_2(z)}]\right)\!: \,, \label{Q_current}\\ 
  &\vec{H}_{f,e}(z) = - :\! \left(e_{i1}(z) e_{i2}(z) \right) \mathop{}\negthickspace \vec{\tau}  \mathop{}\negthickspace \begin{pmatrix} f_{i1}(z) \\ f_{i2}(z) \end{pmatrix} \!:
    - :\! \left(f_{1i}(z) f_{2i}(z) \right) \mathop{}\negthickspace \vec{\tau}  \mathop{}\negthickspace \begin{pmatrix} e_{1i}(z) \\ e_{2i}(z) \end{pmatrix} \!: \,, \nonumber\\\
  &\vec{H}_{\beta, \gamma}(z) = - :\! \left(\gamma_{11j}(z) \gamma_{12j}(z) \right) \mathop{}\negthickspace \vec{\tau}  \mathop{}\negthickspace \begin{pmatrix} \beta_{11j}(z) \\ \beta_{12j}(z) \end{pmatrix} \!: 
  - :\! \left( \beta_{21j}(z) \beta_{22j}(z) \right) \mathop{}\negthickspace \vec{\tau}  \mathop{}\negthickspace \begin{pmatrix} \gamma_{21j}(z) \\ \gamma_{22j}(z) \end{pmatrix} \!: \,. \nonumber\\ \nonumber
  \end{align}
  
  The vacuum is normalized to
  \begin{equation*}
  \bra{0} c_{-1} c_0 c_1 \prod_{k = 1}^3 g_0^i \prod_{i,j=1}^2 \prod_{\alpha, \dot{\alpha} = 1}^2 \left( e^{\dot{\alpha} \alpha}_{ij \, 0} e^{-\varphi^{\dot{\alpha} \alpha}_{1ij}}(0) e^{-\varphi^{\alpha \dot{\alpha}}_{2ij}}(0)\right) \ket{0} = 1 \,.
  \end{equation*}
  The total (ghost,picture) number inside the vacuum normalization is (22,-32). The picture changing operator is
  \begin{align}
  &\chi(z) = \prod_{i,j = 1}^2 \prod_{\alpha, \dot{\alpha} = 1}^2 \chi_{1ij}^{\dot{\alpha} \alpha}(z) \chi_{2ij}^{\dot{\alpha} \alpha}(z),\label{PictureNumber}\\
  &\chi_{1ij}^{\dot{\alpha} \alpha}(z) = \{Q, \xi_{1ij}^{\dot{\alpha} \alpha}(z)\} = c(z)\partial \xi_{1ij}^{\dot{\alpha} \alpha}(z) + e^{\varphi_{1ij}^{\dot{\alpha} \alpha}(z)} G_{1ij}^{\dot{\alpha} \alpha}(z) + \oint \!\! \ud w \vec{g}(w) \{\vec{H}_{\beta, \gamma}(w), \xi_{1ij}^{\dot{\alpha} \alpha}(z)\},\nonumber\\
  &\chi_{2ij}^{\dot{\alpha} \alpha}(z) = \{Q, \xi_{2ij}^{\alpha \dot{\alpha}}(z)\} = c(z)\partial \xi_{2ij}^{\alpha \dot{\alpha}}(z) + e^{\varphi_{2ij}^{\alpha \dot{\alpha}}(z)} G_{2ij}^{\alpha \dot{\alpha}}(z) + \oint \!\! \ud w \vec{g}(w) \{\vec{H}_{\beta, \gamma}(w), \xi_{2ij}^{\alpha \dot{\alpha}}(z)\} \,.
  \nonumber
  \end{align}
  
  Physical states are in the cohomology of the BRST charge. The vacuum normalization shows single occurrence for ghosts with conformal weight 0 such that the Hilbert space needs to include physical states with ghost number 0 for fermionic ghosts except $c$ and picture number 0 for bosonic ghosts\footnote{In section \ref{SFT} the corresponding Hilbert space is denoted by $\widetilde{\mathcal{H}}_1$.}, and such states need to be annihilated by the nonnegative (in particular, the zero) modes of the currents $\vec{H}, F_{ij},  G_{1i j}$, and $G_{2i j}$. This includes states built with only modes from the $\lambda_i, \tilde{\lambda}_j, \phi_k$, and $\tilde{\phi}_l$ fields, with no modes from $\mu_r, \tilde{\mu}_s, \psi_t$, and $\tilde{\psi}_u$. They are (with no implicit summation over repeated indices)
  in the NS sector:
  \begin{equation}
  \begin{aligned}
  &\epsilon_1^{i j \alpha \beta} \lambda_{i \alpha \, -\!\frac{1}{2}} \lambda_{j \beta \, -\frac{1}{2}},
  &&\epsilon_2^{i j \alpha \beta} \lambda_{i \alpha \, -\!\frac{1}{2}} \phi_{j \beta \, -\!\frac{1}{2}},
  &&\epsilon_3^{i j \alpha \dot{\beta}} \lambda_{i \alpha \, -\!\frac{1}{2}} \tilde{\phi}_{j \dot{\beta} \, -\!\frac{1}{2}},
  &&\epsilon_4^{i j \alpha \beta} \phi_{i \alpha \, -\!\frac{1}{2}} \phi_{j\beta \, -\!\frac{1}{2}},\\
  &\epsilon_5^{i j \dot{\alpha} \dot{\beta}} \tilde{\lambda}_{i \dot{\alpha} \, -\!\frac{1}{2}} \tilde{\lambda}_{j \dot{\beta} \, -\!\frac{1}{2}},
  &&\epsilon_6^{i j \dot{\alpha} \dot{\beta}} \tilde{\lambda}_{i \dot{\alpha} \, -\!\frac{1}{2}} \tilde{\phi}_{j \dot{\beta} \, -\!\frac{1}{2}},
  &&\epsilon_7^{i j \dot{\alpha} \beta} \tilde{\lambda}_{i \dot{\alpha} \, -\!\frac{1}{2}} \phi_{j \beta \, -\!\frac{1}{2}},
  &&\epsilon_8^{i j \dot{\alpha} \dot{\beta}} \tilde{\phi}_{i \dot{\alpha} \, -\!\frac{1}{2}} \tilde{\phi}_{j \dot{\beta} \, -\!\frac{1}{2}},\\
  &\epsilon_9^{i j \alpha \dot{\beta}} \lambda_{i \alpha \, -\!\frac{1}{2}} \tilde{\lambda}_{j \dot{\beta} \, -\!\frac{1}{2}},
  &&\epsilon_{10}^{i j \alpha \dot{\beta}} \phi_{i \alpha \, -\!\frac{1}{2}} \tilde{\phi}_{j  \dot{\beta} \, -\frac{1}{2}},
  \end{aligned}
  \label{NSSpectrum}
  \end{equation}
  where the $\epsilon_k$ are constants, and in the R sector:
  \begin{equation*}
  \begin{aligned}
  &[\tilde{\lambda}_{i\, -1} f_i(\cdots)],
  &&\braket{\lambda_{i \, -1} g_i(\cdots)},
  &[\tilde{\phi}_{i  \, -1} \mathcal{f}(\cdots)],
  &&\braket{\phi_{i \, -1} \mathcal{g}(\cdots)},\\
  \end{aligned}
  \end{equation*}
  where $f_i^{\dot{\alpha}}, g_i^{\alpha}, \mathcal{f}^{\dot{\alpha}}$, and $\mathcal{g}^{\alpha}$ are functions of zero modes $\lambda_{j \beta \, 0}, \mu_{k \, 0}^{\dot{\beta}}, \phi_{r \delta \, 0}$, and $\psi_{s \, 0}^{\dot{\delta}}$, with $\mu_{k \, 0}^{\dot{\beta}}$ and $\psi_{s \, 0}^{\dot{\delta}}$ ultimately set to zero or reduced to $\lambda_{j \beta \, 0}$ and $\phi_{r \delta \, 0}$ using gauge transformations. The functions must be chosen such that
  \begin{equation*}
  \begin{aligned}
  &[\frac{\partial}{\partial \mu_{i \, 0}} f_i(\cdots)] = 0 , && \braket{\lambda_{i \, 0} g_i(\cdots)} = 0,
  &[\frac{\partial}{\partial \psi_{i \, 0}} \mathcal{f}(\cdots)] = 0 , && \braket{\phi_{i \, 0} \mathcal{g}(\cdots)} = 0,
  \end{aligned}
  \end{equation*}
  and additionally the functions $f_i$ and $\mathcal{f}$ have to be homogeneous of degree 1 and $g_i$ and $\mathcal{g}$ homogeneous of degree -1. The first and third condition seem unnecessary by using functions of $\lambda_{j \beta \, 0}$ and $\phi_{r \delta \, 0}$ only, but in section \ref{Spectrum}, where we use a similar form to define fixed vertex operators but in ghost number 1 for every fermionic ghost and picture number $-1$ for every bosonic ghost, we allow functions $f_i^{\dot{\alpha}}, g_i^{\alpha}, \mathcal{f}^{\dot{\alpha}}$, and $\mathcal{g}^{\alpha}$ to be functions of all matter fields, and then they need to satisfy all four conditions.
  
  Finally, in both sectors the states have to be gauge-invariant with respect to the SU(2) symmetry of the bosonic twistors, i.e. annihilated by $\vec{H}_n$ for $n \ge 0$. E.g. $\sum_i \tilde{\lambda}_{i \dot{\alpha} \, -1} f_i^{\dot{\alpha}}$ can be chosen as an SU(2) singlet if $\tilde{\lambda}_i$ and $f_i$ transform inverse under SU(2) transformations. In the R sector there can be many possible SU(2) representations because the number of zero mode twistors is not predetermined.
  
  Because our target space includes the dual twistor space in a symmetric fashion, we have to add the dual physical states in the R sector, but in order to do that we have to double the Hilbert space in the R sector and in the new half switch the zero mode matter operators that annihilate the vacuum to the dual ones:
 \begin{align*}
 \lambda_{i\,n\!-\!\frac{1}{2}} \ket{0} &= 0, &\mu_{i \, n\!-\!\frac{1}{2}} \ket{0} &= 0, &\tilde{\lambda}_{i\,n}  \ket{0} &= 0, &\tilde{\mu}_{i \, n} \ket{0} &= 0, &\\
 \phi_{i\,n\!-\!\frac{1}{2}} \ket{0} &= 0, &\psi_{i \, n\!-\!\frac{1}{2}} \ket{0}  &= 0, &\tilde{\phi}_{i\,n}  \ket{0} &= 0, &\tilde{\psi}_{i \, n} \ket{0} &= 0, &
  \end{align*}
  for $n > 0$. Physical states in the BRST cohomology residing in the dual part of the Hilbert space are now:
  \begin{equation*}
  \begin{aligned}
  &\braket{\lambda_{i  \, -1} \tilde{f}_i(\cdots)},
  &&[\tilde{\lambda}_{i \, -1} \tilde{g}_i(\cdots)],
  &&\braket{\phi_{i \, -1} \tilde{\mathcal{f}}(\cdots)}, 
  &&[\tilde{\phi}_{i \, -1} \tilde{\mathcal{g}}(\cdots)],
   \end{aligned}
  \end{equation*}
  where $\tilde{f}_i^{\alpha}, \tilde{g}_i^{\dot{\alpha}}, \tilde{\mathcal{f}}^{\alpha}$, and $\tilde{\mathcal{g}}^{\dot{\alpha}}$ are functions of $\tilde{\lambda}_{j \dot{\beta} \, 0}, \tilde{\mu}_{k \, 0}^{\beta}, \tilde{\phi}_{r \dot{\delta} \, 0},$ and $\tilde{\psi}_{s \, 0}^{\delta}$, with $\tilde{\mu}_{k \, 0}^{\beta}$ and $\tilde{\psi}_{s \, 0}^{\delta}$ ultimately set to zero or reduced to $\tilde{\lambda}_{j \dot{\beta} \, 0}$ and $\tilde{\phi}_{r \dot{\delta} \, 0}$ using gauge transformations, and must be chosen such that
  \begin{equation*}
  \begin{aligned}
  &\braket{\frac{\partial}{\partial \tilde{\mu}_{i \, 0}} \tilde{f}_i(\cdots)} = 0, &&[\tilde{\lambda}_{i \, 0} \tilde{g}_i(\cdots)] = 0,
  &\braket{\frac{\partial}{\partial \tilde{\psi}_{i \, 0}} \tilde{\mathcal{f}}(\cdots)} = 0, &&[\tilde{\phi}_{i \, 0} \tilde{\mathcal{g}}(\cdots) = 0,\\
  \end{aligned}
  \end{equation*}
  and, of course, additionally the states have to be gauge-invariant with respect to SU(2), $\tilde{f}_i$ and $\tilde{\mathcal{f}}$ homogenous of degree 1, and $\tilde{g}_i$ and $\tilde{\mathcal{g}}$ homogeneous of degree -1.\\
    
  Notice that in \cite{Kunz:2020} the assignment of the zero modes annihilating the vacuum was different: $\lambda_{i 0} \ket{0} = \tilde{\lambda}_{j 0} \ket{0} = 0$. This poses a problem for a particle interpretation using plane waves with the zero modes providing the momentum, i.e. the vertex operators in section \ref{Spectrum} for the R sector would not be usable.\\

\section{Four-Point One-Loop Graviton Amplitude from Single Cut} 
\label{4GravOneLoop}
Here it will be shown that the four-graviton one-loop MHV amplitude in the non-separating factorization limit leads to the expected result found in the literature. For this we use the solution set for the 6-particle scattering equations with equal number of positive and negative helicities provided in appendix C of \cite{Farrow:2020}, but with zero BCFW shift. The particles are partitioned into an $L=\{1,2\}$ set of negative helicities, an $R=\{3,4\}$ set of positive helicities, and two auxiliary punctures $\{-,+\}$ for the loop particle, with scattering equations
\begin{equation*}
\begin{aligned}
&\lambda_0 = \sum_{l \in L} \frac{\lambda_l}{(+l)}, &\tilde{\lambda}_0 = \sum_{r \in R} \frac{\tilde{\lambda}_r}{(-r)}
\end{aligned}
\end{equation*}
for the auxiliary punctures, and
\begin{equation*}
\begin{aligned}
&\lambda_r = \sum_{l \in L} \frac{\lambda_l}{(rl)} + \frac{\lambda_0}{(r-)}, &\tilde{\lambda}_l = \sum_{r \in R} \frac{\tilde{\lambda}_r}{(lr)} -  \frac{\tilde{\lambda}_0}{(l+)}
\end{aligned}
\end{equation*}
for the remaining particles.
The solutions in terms of homogeneous coordinates $\sigma$, after gauge fixing particles 1 and 2 to $\sigma = (1,0)$ and $\sigma = (0,1)$, respectively, are:
\begin{equation*}
\sigma =
\begin{bmatrix}
1 & 0 & \frac{[34] \!\braket{12}}{\bra{1}3+0|4]}& -\frac{[34] \!\braket{12}}{\bra{1}4+0|3]} & \frac{\braket{12}}{\braket{10}} & \frac{\bra{2}3+0|4]\bra{2}4+0|3][20]}{P^2_{034} \braket{12}[03][04]}\\
0 & 1 & \frac{[34] \!\braket{12}}{\bra{2}3+0|4]}& -\frac{[34] \!\braket{12}}{\bra{2}4+0|3]} & \frac{\braket{12}}{\braket{20}} & \frac{\bra{1}3+0|4]\bra{1}4+0|3][01]}{P^2_{034} \braket{12}[03][04]}
\end{bmatrix} \,,
\end{equation*}
where the columns are labeled 1,2,3,4,+,-, $P_{034} = \frac{1}{\sqrt{2}}(\lambda_0 \tilde{\lambda}_0 + \lambda_3 \tilde{\lambda}_3 + \lambda_4 \tilde{\lambda}_4$), and the notation $\bra{i}j+0|k] = \braket{ij}[jk] + \braket{i0}[0k]$ has been used. Then the scattering amplitude \eqref{single_cut} is
\begin{equation*}
\mathcal{M}_{2,4}^{1-loop} \!= \delta^4(P) \int \frac{\ud^2 \!\lambda_0  \ud^2 \!\tilde{\lambda}_0}{\mathrm{vol} \mathbb{C}^{*}} \sum_{i \in L \atop{j \in R} } \lim_{(l +) \to  \sigma_l^1 \sigma_+^2 \atop{(\forall l \in L)}} \lim_{(l -) \to  \sigma_l^1 \sigma_-^2 \atop{(\forall l \in R)}} \mathcal{J} \frac{\braket{12}[34]}{(12)(34)}\frac{\braket{0 \, i}}{(- \, i)} \frac{[0 \, j]}{(+ \, j)} \,,
\end{equation*}
where the Jacobian $\mathcal{J}$ is
\begin{equation*}
\mathcal{J} = \frac{[34]^8 \braket{12}^8}{P_{034}^2 \bra{1}3+0|4] \bra{1}4+0|3]\bra{2}3+0|4] \bra{2}4+0|3] [03]^2 [04]^2 \braket{01}^2 \braket{02}^2} \,.
\end{equation*}
The part of the integrand not involving contractions with the loop particle combines simply to:
\begin{equation*}
\mathcal{J} \frac{\braket{12}[34]}{(12)(34)} = \frac{[34]^6 \braket{12}^6}{(P_{034}^2)^2 [03]^2 [04]^2 \braket{01}^2 \braket{02}^2} \,.
\end{equation*}
For the rest of the integrand we apply the limit $(4 -) \to  \sigma_4^1 \sigma_-^2$:
\begin{equation*}
(4 \, -) = \frac{[3 \, 4]}{[0 \, 3]} \cong \sigma_4^1 \sigma_-^2 = - \frac{[3 \, 4] [0 \, 1] \bra{1}3+0|4]} {P_{034}^2 [0 \, 3] [0 \, 4]}
\end{equation*}
and use permutations $(3 \leftharpoondown \!\!\!\!\!\! \rightharpoonup 4)$ and $(1 \leftharpoondown \!\!\!\!\!\! \rightharpoonup 2)$\footnote{The limits $(l +) \to  \sigma_l^1 \sigma_+^2 (l = 1,2)$ are not useful here because of gauge fixing of particles 1 and 2.}, leading to
\begin{equation*}
\begin{aligned}
& \bra{1}3+0|4] \cong - \frac{P_{034}^2 [0 \, 4]}{[0 \, 1]}, &\bra{1}4+0|3] \cong - \frac{P_{034}^2 [0 \, 3]}{[0 \, 1]},\\
& \bra{2}3+0|4] \cong - \frac{P_{034}^2 [0 \, 4]}{[0 \, 2]}, &\bra{2}4+0|3] \cong - \frac{P_{034}^2 [0 \, 3]}{[0 \, 2]}\,.
\end{aligned}
\end{equation*}
Then 
\begin{equation*}
\begin{split}
&(- \, 1) = - \frac{\bra{1}3+0|4] \bra{1}4+0|3] [0 \,1]}{P^2_{034} \braket{12}[03][04]} \cong - \frac{P^2_{034}}{\braket{1 \, 2} [0 \,1]} \,,\\
&(+ \, 3) = - \frac{[3 \, 4]^2 \braket{1 \, 2}^3 \braket{0 \, 3}}{\bra{2}3+0|4] \bra{1}3+0|4] \braket{1 \, 0} \braket{2 \, 0}} \cong - \frac {[3 \, 4]^2 \braket{1 \, 2}^3 \braket{0 \, 3} [0 \, 1] [0 \, 2]}{(P^2_{034})^2 [0 \, 4]^2 \braket{0 \, 1} \braket{0 \, 2}} \,,
\end{split}
\end{equation*}
with similar equations for $(- \, 2)$ and $(+ \, 4)$ and the amplitude becomes
\begin{equation*}
\begin{split}
\mathcal{M}_{2,4}^{1-loop} \!&= \delta^4(P) \int \frac{\ud^2 \!\lambda_0  \ud^2 \!\tilde{\lambda}_0}{\mathrm{vol} \mathbb{C}^{*}} \sum_{i \in L \atop{j \in R} } \frac{[3 \, 4]^4 \braket{1 \, 2}^4}{P_{034}^2 \braket{0 \, i} [i \, 0] \braket{0 \, j} [j \, 0]} \\
&\cong \delta^4(P) \int \ud^4 \ell \frac{[3 \, 4]^4 \braket{1 \, 2}^4}{\ell^2 (\ell + k_1)^2 (\ell + k_1 + k_2)^2 (\ell - k_3)^2} + \,\text{perm}(1 \leftharpoondown \!\!\!\!\!\! \rightharpoonup 2, 3 \leftharpoondown \!\!\!\!\!\! \rightharpoonup 4) \,.
\end{split}
\end{equation*}
This is $[3 \, 4]^2 \! \braket{1 \, 2}^{\!2}$ times the result for the one-loop $\mathcal{N}=4$ SYM amplitude obtained in appendix C of \cite{Farrow:2020} and is the expected result \cite{Bern:1998}\footnote{The same relation between the two amplitudes was also obtained in section 5 of \cite{Farrow:2017}.}.\\

\appendix

\bibliography{TwistorString}
\end{document}